\documentclass[aps,prc,twocolumn,preprintnumbers,amsmath,
amssymb,superscriptaddress,nofootinbib]{revtex4-2} 
\usepackage{dcolumn}
\usepackage{multirow}
\usepackage{bm}
\usepackage{graphicx}
\usepackage{textcomp}
\usepackage{graphicx}
\usepackage{longtable}
\usepackage{booktabs}
\usepackage{ulem}
\usepackage[colorlinks=true,linkcolor=blue,citecolor=blue,urlcolor=blue]{hyperref}
\usepackage[T1]{fontenc}
\usepackage{xcolor}
\graphicspath{{Images/}}

\newcolumntype{C}{>{$}c<{$}}
\AtBeginDocument{
	\heavyrulewidth=.08em
	\lightrulewidth=.05em
	\cmidrulewidth=.03em
	\belowrulesep=.65ex
	\belowbottomsep=0pt
	\aboverulesep=.4ex
	\abovetopsep=0pt
	\cmidrulesep=\doublerulesep
	\cmidrulekern=.5em
	\defaultaddspace=.5em
}

\begin{document}
	
	\newcommand{\ts}{\textsuperscript}
	\newcommand{\pieter}{\textcolor{red}}
	\newcommand{\linh}{\textcolor{purple}}
	\newcommand{\anna}{\textcolor{blue}}
	\newcommand{\nancy}{\textcolor{orange}}
	\newcommand{\corref}{\textcolor{green}}
	\newcommand{\annabis}{\textcolor{magenta}}
	
	\title{Investigation of the ground-state spin inversion in the neutron-rich \texorpdfstring{$^{47,49}$Cl}{Lg} isotopes}
	
	\newcommand{\acea}{        \affiliation{Universit\'e Paris-Saclay, IRFU, CEA, F-91191 Gif-sur-Yvette, France}}
	\newcommand{\ariken}{      \affiliation{RIKEN Nishina Center, 2-1 Hirosawa, Wako, Saitama 351-0198, Japan}}
	\newcommand{\atud}{	       \affiliation{Institut f\"ur Kernphysik, Technische Universit\"at Darmstadt, 64289 Darmstadt, Germany}}
	\newcommand{\aiphc}{       \affiliation{IPHC, CNRS/IN2P3, Universit\'e de Strasbourg, F-67037 Strasbourg, France}}
	\newcommand{\acns}{        \affiliation{Center for Nuclear Study, University of Tokyo, RIKEN campus, Wako, Saitama 351-0198, Japan}}
	\newcommand{\aut}{         \affiliation{Department of Physics, University of Tokyo, 7-3-1 Hongo, Bunkyo, Tokyo 113-0033, Japan}}
	\newcommand{\arikkyo}{     \affiliation{Department of Physics, Rikkyo University, 3-34-1 Nishi-Ikebukuro, Toshima, Tokyo 172-8501, Japan}}
	\newcommand{\abeijing}{    \affiliation{State Key Laboratory of Nuclear Physics and Technology, Peking University, Beijing 100871, P.R. China}}
	\newcommand{\ainst}{    \affiliation{Institute for Nuclear Science \& Technology, VINATOM, 179 Hoang Quoc Viet, Cau Giay, Hanoi, Vietnam}}
	\newcommand{\aatomki}{     \affiliation{Institute for Nuclear Research, Atomki, P.O. Box 51, Debrecen H-4001, Hungary}}
	\newcommand{\aipno}{       \affiliation{Institut de Physique Nucl\'eaire Orsay, IN2P3-CNRS, 91406 Orsay Cedex, France}}
	\newcommand{\aijclab}{       \affiliation{Universit\'e Paris-Saclay, CNRS/IN2P3, IJCLab, Orsay, France}}
	\newcommand{\aus}{       \affiliation{Departamento de Fisica Atomica Molecular y Nuclear, Facultad de Fisica, Universidad de Sevilla, Apartado 1065, E-41080 Sevilla, Spain}}
	\newcommand{\aoslo}{       \affiliation{Department of Physics, University of Oslo, N-0316 Oslo, Norway}}
	\newcommand{\ahku}{        \affiliation{Department of Physics, The University of Hong Kong, Pokfulam, Hong Kong}}
	\newcommand{\atohoku}{     \affiliation{Department of Physics, Tohoku University, Sendai 980-8578, Japan}}
	\newcommand{\acolb}{	     \affiliation{Pontificia Universidad Javeriana, Facultad de Ciencias, Departamento de F\'isica, Bogot\'a, Colombia}}
	\newcommand{\acol}{	     \affiliation{Universidad Nacional de Colombia, Sede Bogot\'a, Facultad de Ciencias, Departamento de F\'isica, Bogot\'a 111321, Colombia}}
	\newcommand{\arik}{        \affiliation{Department of Physics, Rikkyo University, 3-34-1 Nishi-Ikebukuro, Toshima, Tokyo 172-8501, Japan}}
	\newcommand{\astock}{       \affiliation{Department of Physics, Royal Institute of Technology, SE-10691 Stockholm, Sweden}}
	\newcommand{\agsi}{       \affiliation{GSI Helmholtzzentrum f\"ur Schwerionenforschung GmbH, Planckstr. 1, 64291 Darmstadt, Germany}}
	\newcommand{\akoln}{       \affiliation{Institut f\"{u}r Kernphysik, Universit\"{a}t zu K\"{o}ln, D-50937 K\"{o}ln, Germany}}
	\newcommand{\alpc}{       \affiliation{LPC Caen, Normandie Univ, ENSICAEN, UNICAEN, CNRS/IN2P3, F-14000 Caen, France}}
	\newcommand{\akorea}{       \affiliation{Ewha Womans University, Seoul 03760, Korea}}
	\newcommand{\aibs}{\affiliation{Institute for Basic Science, Daejeon 34126, Korea}}
	\newcommand{\atit}{     \affiliation{Department of Physics, Tokyo Institute of Technology, 2-12-1 O-Okayama, Meguro, Tokyo 152-8551, Japan}}
	\newcommand{\amad}{       \affiliation{Instituto de Estructura de la Materia, CSIC, E-28006 Madrid, Spain}}
	\newcommand{\alanz}{       \affiliation{Institute of Modern Physics, Chinese Academy of Sciences, Lanzhou 730000, China}}
	
	\newcommand{\amiuni}{        \affiliation{Dipartimento di Fisica, Universit\`a degli Studi di Milano, Via Celoria 16, I-20133 Milano, Italy}}
	\newcommand{\amiinfn}{        \affiliation{INFN, Sezione di Milano, Via Celoria 16, I-20133 Milano, Italy}}
	\newcommand{\aleu}{        \affiliation{KU Leuven, Instituut voor Kern- en Stralingsfysica, B-3001 Leuven, Belgium}}
	\newcommand{\atriumf}{        \affiliation{TRIUMF, 4004 Wesbrook Mall, Vancouver, British Columbia, Canada V6T 2A3}}
	\newcommand{\amcgill}{ 
		\affiliation{Department of Physics, McGill University, 3600 Rue University, Montr\'eal, QC H3A 2T8, Canada}}
	\newcommand{\asurrey}{     \affiliation{Department of Physics, University of Surrey, Guildford GU2 7XH, United Kingdom}}
	\newcommand{\aosaka}{     \affiliation{Research Center for Nuclear Physics (RCNP), Osaka University, Ibaraki 567-0047, Japan}}
	\newcommand{\aosakab}{     \affiliation{Department of Physics, Osaka City University, Osaka 558-8585, Japan}}
	\newcommand{\asaigon}{     \affiliation{Department of Nuclear Physics, Faculty of Physics and Engineering
			Physics, University of Science, Ho Chi Minh City, Vietnam}}
	\newcommand{\asaigonb}{     \affiliation{Vietnam National University, Ho Chi Minh City, Vietnam}}
	\newcommand{\ajaea}{     \affiliation{Japan Atomic Energy Agency, Tokai, Ibaraki 319-1195, Japan}}
	\newcommand{\aviet}{     \affiliation{Vietnam Agency for Radiation and Nuclear Safety, 113 Tran Duy Hung, Cau Giay, Hanoi, Vietnam}}
	\newcommand{\abro}{     \affiliation{Laboratoire Kastler Brossel, Sorbonne Universit\'e, CNRS, ENS, PSL Research University, Coll\`ege de France,
			Case 74, 4 Place Jussieu, 75005 Paris, France}}
	\newcommand{\azagreb}{     \affiliation{Ru{\dj}er Bo\v{s}kovi\'c Institute, Bijeni\v{c}ka cesta 54, 10000 Zagreb, Croatia}}
	
	\author{B.D.~Linh}      \ainst
	\author{A.~Corsi}        \acea
	\author{A.~Gillibert}    \acea
	\author{A.~Obertelli}     \acea \ariken \atud
	\author{P.~Doornenbal}    \ariken
	\author{C.~Barbieri}         \amiuni \amiinfn \asurrey
	\author{S.~Chen}       \ahku \ariken \abeijing
	\author{L.X.~Chung}      \ainst
	\author{T.~Duguet}       \acea \aleu     
	\author{M.~G\'omez-Ramos}     \atud \aus
	\author{J.D.~Holt}       \atriumf  \amcgill
	\author{A.~Moro}     \aus
	\author{P.~Navr\'atil}  \atriumf      
	\author{K.~Ogata}  \aosaka \aosakab
	\author{N.T.T.~Phuc}  \asaigon \asaigonb 
	\author{N.~Shimizu}       \acns      
	\author{V.~Som\`a}       \acea      
	\author{Y.~Utsuno}           \acns   \ajaea    
	\author{N. L.~Achouri}       \alpc
	\author{H.~Baba}         \ariken
	\author{F.~Browne}         \ariken
	\author{D.~Calvet}       \acea
	\author{F.~Ch\^ateau}    \acea
	\author{N.~Chiga}       \ariken
	\author{M. L.~Cort\'es}       \ariken 
	\author{A.~Delbart}      \acea
	\author{J.-M.~Gheller}   \acea
	\author{A.~Giganon}      \acea
	\author{C.~Hilaire}      \acea
	\author{T.~Isobe}        \ariken
	\author{T.~Kobayashi}        \atohoku
	\author{Y.~Kubota}        \ariken \acns
	\author{V.~Lapoux}       \acea
	\author{H. N.~Liu}       \acea \atud \astock
	\author{T.~Motobayashi}  \ariken      
	\author{I.~Murray}   \aijclab \ariken     
	\author{H.~Otsu}         \ariken
	\author{V.~Panin}      \ariken
	\author{N.~Paul}      \acea \abro
	\author{W.~Rodriguez}  \ariken  \acolb   \acol
	\author{H.~Sakurai}      \ariken \aut
	\author{M.~Sasano}       \ariken      
	\author{D.~Steppenbeck}   \ariken
	\author{L.~Stuhl}        \acns  \aatomki \aibs 
	\author{Y. L.~Sun}      \acea \atud  
	\author{Y.~Togano}     \arik
	\author{T.~Uesaka}       \ariken      
	\author{K.~Wimmer}       \aut \ariken
	\author{K.~Yoneda}       \ariken
	\author{O.~Aktas}       \astock
	\author{T.~Aumann}  \atud \agsi
	\author{F.~Flavigny}     \aijclab \alpc 
	\author{S.~Franchoo}     \aijclab 
	\author{I.~Ga\v{s}pari\'c}     \azagreb \atud \ariken
	\author{R.-B.~Gerst}     \akoln
	\author{J.~Gibelin}	\alpc
	\author{K.I.~Hahn}      \akorea \aibs
	\author{N.T.~Khai}      \aviet
	\author{D.~Kim}      \akorea \ariken \aibs
	\author{T.~Koiwai}        \aut
	\author{Y.~Kondo}       \atit
	\author{P.~Koseoglou}     \atud \agsi
	\author{J.~Lee}          \ahku
	\author{C.~Lehr}     \atud
	\author{T.~Lokotko}       \ahku
	\author{M.~MacCormick}       \aijclab 
	\author{K.~Moschner}     \akoln
	\author{T.~Nakamura}    \atit      
	\author{S.Y.~Park}      \akorea \aibs
	\author{D.~Rossi}          \atud
	\author{E.~Sahin}        \aoslo
	\author{D.~Sohler}        \aatomki
	\author{P.-A.~S\"oderstr\"om}	\atud
	\author{S.~Takeuchi}     \atit  
	\author{N.D.~Ton}    \ainst
	\author{H.~T\"ornqvist}  \atud \agsi
	\author{V.~Vaquero}     \amad
	\author{V.~Wagner}       \atud
	\author{H.~Wang}         \alanz
	\author{V.~Werner}       \atud
	\author{X.~Xu}           \ahku 
	\author{Y.~Yamada}       \atit
	\author{D.~Yan}           \alanz
	\author{Z.~Yang}    \ariken
	\author{M.~Yasuda}       \atit
	\author{L.~Zanetti}           \atud

	\begin{abstract}
		A first $\gamma$-ray study of $^{47,49}$Cl spectroscopy was performed at the Radioactive Isotope Beam Factory with $^{50}$Ar projectiles at 217 MeV/nucleon, impinging on the liquid hydrogen target of the MINOS device. Prompt de-excitation $\gamma$ rays were measured with the NaI(Tl) array DALI2\ts{+}. Through the one-proton knockout reaction $^{50}$Ar(p,2p), a spin assignment could be determined for the low-lying states of $^{49}$Cl from the momentum distribution obtained with the SAMURAI spectrometer. A spin-parity $J^{\pi} = 3/2^{+}$ is deduced for the ground state of $^{49}$Cl, similar to the recently studied $N=32$ isotope $^{51}$K. The evolution of the energy difference $E(1/2^{+}_{1}) - E (3/2^{+}_{1})$ is compared to state-of-the-art theoretical predictions.
	\end{abstract}
	
	
	\maketitle
	\section{Introduction}
	In the simplest shell-model framework, nucleons inside nuclei may be considered as independent particles, subject only to the mean-field potential created by the other nucleons. From systematics of the first $2^{+}$ excited state energies, $E(2^{+}_{1})$, for even-even neutron-rich nuclei in the vicinity of calcium isotopes, a strong shell effect is visible at the well-established magic number $N = 28$ with a maximum of $E(2^{+}_{1})$ for the Ca, Ar and S isotopic series. Not only the $E(2^{+}_{1})$ energies but also other quantities, including masses and reduced transition probabilities $B(E2; 0^{+}_{1} \rightarrow 2^{+}_{1}$), are relevant observables for understanding these shell effects. It has been shown that far from stability, the usual ordering of nuclear shells and energy gaps evolves, resulting in new magic numbers such as $N=32, 34$ in Ar and Ca isotopes~\cite{liu,huck,gad,wien,step}, while known ones may disappear, such as $N =28$ for Si, Mg~\cite{bas,craw}. More data extended to the neighboring isotopic series will characterize the evolution of these shell effects. 
	
	In parallel to the modification of the neutron $pf$ shell structure, the addition of neutrons in the Ca isotopes modifies the ordering of the proton orbitals. Valuable information on this ordering can be provided by the spectroscopy of odd-$Z$ nuclei. In the K isotopes, a naive expectation from the shell model gives $J^{\pi} = 3/2^+$ for the ground state and $J^{\pi} = 1/2^+$ for the first excited state, corresponding to a proton hole in the $0d_{3/2}$ and $1s_{1/2}$ orbitals, respectively, as observed for the stable nucleus $^{39}$K~\cite{dol1}. The potassium isotopic series has been well documented through transfer~\cite{ynt,dol1,ogi}, $\beta$-decay~\cite{huck,bro,wei}, and laser-spectroscopy studies~\cite{papu1,papu2}. As the ground-state spin parity was unambiguously assigned, the low-lying level spectroscopy can be studied with a variation of the first excited state energy with the neutron number $N$. The energy for the $1/2^+_1$ level decreases when neutrons are added from $N = 20$ to 28 and increases beyond $N$=28~\cite{papu1}. The maximum effect is observed for $^{47}$K at $N = 28$ with spin inversion and a $J^{\pi} = 1/2^{+}$ ground state dominated by the $\pi(1s_{1/2})^{-1} \otimes \nu$(pf) configuration. Spin inversion is still observed for $^{49}$K, but a $3/2^{+}$ ground state is restored for $^{51}$K at $N = 32$. Recently, the first spectroscopy of $^{51,53}$K was performed~\cite{sun} with a $3/2^{+}$ ground state for $^{53}$K and a first excited state assigned as $1/2^{+}$. Experimental g factors were also determined for the ground state~\cite{papu1} and are consistent with effective values calculated for a proton hole in the $\pi 0d_{3/2}$ orbital for $N < 28$ and $N = 32$, and a proton hole in the $\pi 1s_{1/2}$ orbital for $N=28$.
	
	The reduction of the energy difference $\Delta =[E(1/2^{+}_1) - E(3/2^{+}_1)]$ with increasing neutron number for the $N \leq 28$ potassium isotopes has been interpreted as being due to the tensor interaction between the $\pi 0d_{3/2}$ ($j_< \equiv l - 1/2$) and $\nu 0f_{7/2}$ ($j_> \equiv l + 1/2$) orbitals, with $\Delta J = 2$, while the $\pi 1s_{1/2}$ orbital is unaffected. Filling the $\nu 0f_{7/2}$ orbital from $N = 20$ to 28 has an increasing attractive effect on the $\pi 0d_{3/2}$ orbital~\cite{otsu2}: the gap between the $\pi 0d_{3/2}$ and $\pi 1s_{1/2}$ orbitals decreases until the $\pi 1s_{1/2}$ orbital becomes the valence orbital for $N=28$. 
	
	Increasing collectivity is expected when moving away from the closed-shell Ca core. With two fewer protons, the Ar isotopes are well suited for the study of collective effects and can be compared to the calcium isotones. A similar comparison may be performed on the odd partners, potassium and chlorine isotopes. With (d,$^{3}$He) transfer experiments on even argon isotopes~\cite{doll,mair}, the $3/2^{+}_{1}$ ground state and $1/2^{+}_{1}$ first excited state were identified in $^{37,39}$Cl as proton-hole states with large spectroscopic factors exhausting most of the corresponding strength. A sharp decrease was observed for the energy difference $ \Delta$ from $^{37}$Cl to $^{39}$Cl, suggesting a possible spin inversion and a $1/2^{+}_{1}$ ground state for more exotic isotopes.
	
	Beyond $N = 22$, a possible spin inversion is also predicted in theoretical calculations~\cite{sor}, with a very small energy difference $ \Delta$ for
	$^{41,43,45}$Cl. These isotopes were studied with 
	$\beta$-decay and in-beam $\gamma$-ray spectroscopy using various reaction mechanisms.
	Since no spin parity measurement could be performed, only the absolute value $\vert \Delta \vert$ was determined. For $^{41}$Cl, a small value $\vert \Delta \vert = 129.7 \text{ keV}$ was found in in-beam studies using deep-inelastic scattering~\cite{lia,oll,szi}, and is consistent with shell-model calculations. $\beta$-decay data~\cite{wiz} were inconclusive on the spin assignment. Similar information is available for $^{43,45}$Cl with $\vert \Delta \vert$ = 328 and 127~keV. First spectroscopy studies for $^{43,45}$Cl~\cite{ibb,gad,sor} were performed under the assumption of a $1/2^{+}_{1}$ ground state, as predicted in the shell-model calculation~\cite{sor}. However, further studies seem to be in contradiction with this first hypothesis~\cite{win,stro,ril}, calling into question the former spin parity assignment for the ground state of $^{43,45}$Cl.
	
	Additional information was provided by the measurement of the g factor for $^{44}$Cl, which was consistent with a 2$^{-}$ ground state~\cite{rydt}. The shell model calculations performed at $N = 27$ for K, Cl isotopes conclude a dominant $\pi 0d_{3/2}$ hole configuration for the ground state of $^{46}$K, while configuration mixing is present in $^{44}$Cl with a sizable $\pi 1s_{1/2}$ component in the ground state wave function. Due to the three proton holes in the closed $Z=20$ proton core, odd-$Z$ chlorine isotopes are expected to be sensitive to the relative position of the $\pi 0d_{3/2}$ and $\pi 1s_{1/2}$ orbitals, which are strongly impacted by the neutron number and the filling of the $\nu 0f_{7/2}$ orbitals and beyond.
	
	We report on the first spectroscopy of neutron-rich $^{47,49}$Cl isotopes and the investigation of the ground-state spin inversion in Cl isotopes far from stability. Data are compared to state-of-the-art theoretical calculations, including shell model and \textit{ab initio} methods. These latter methods are now able to compute open shell, intermediate-mass, odd-even nuclei in their full-space implementation~\cite{soma11, soma20b}, and all nuclei accessible to standard shell-model approaches via valence-space techniques~\cite{stroberg19}. Neutron-rich nuclei constitute an important benchmark for the development of both many-body methods and input nuclear Hamiltonians, currently modeled within the framework of chiral effective field theory (EFT)~\cite{epelbaum09, machleidt11}. In the present work, new \textit{ab initio} calculations were performed within the valence-space formulation of the in-medium similarity renormalization group (VS-IMSRG)~\cite{bogner14, stroberg16,stroberg19,simonis16} and the Gorkov self-consistent Green's function (GGF)~\cite{soma11, soma14a} approaches.
	
	The experimental setup is described in Section II, and the methods used for data analysis are developed in Section III, including the determination of momentum distributions and cross sections. The experimental results for $^{49}$Cl and $^{47}$Cl are detailed in Sections IV and V, respectively. In Section VI, data are compared to the shell model predictions with phenomenological SDPF-MU interaction and VS-IMSRG-derived interactions (called SDPF-MU calculation and IMSRG calculation in the following), and full-space calculations performed within the GGF approach.
	
	\begin{figure}[hbtp!]
		\begin{center}			
		\includegraphics[width=8.6 cm]{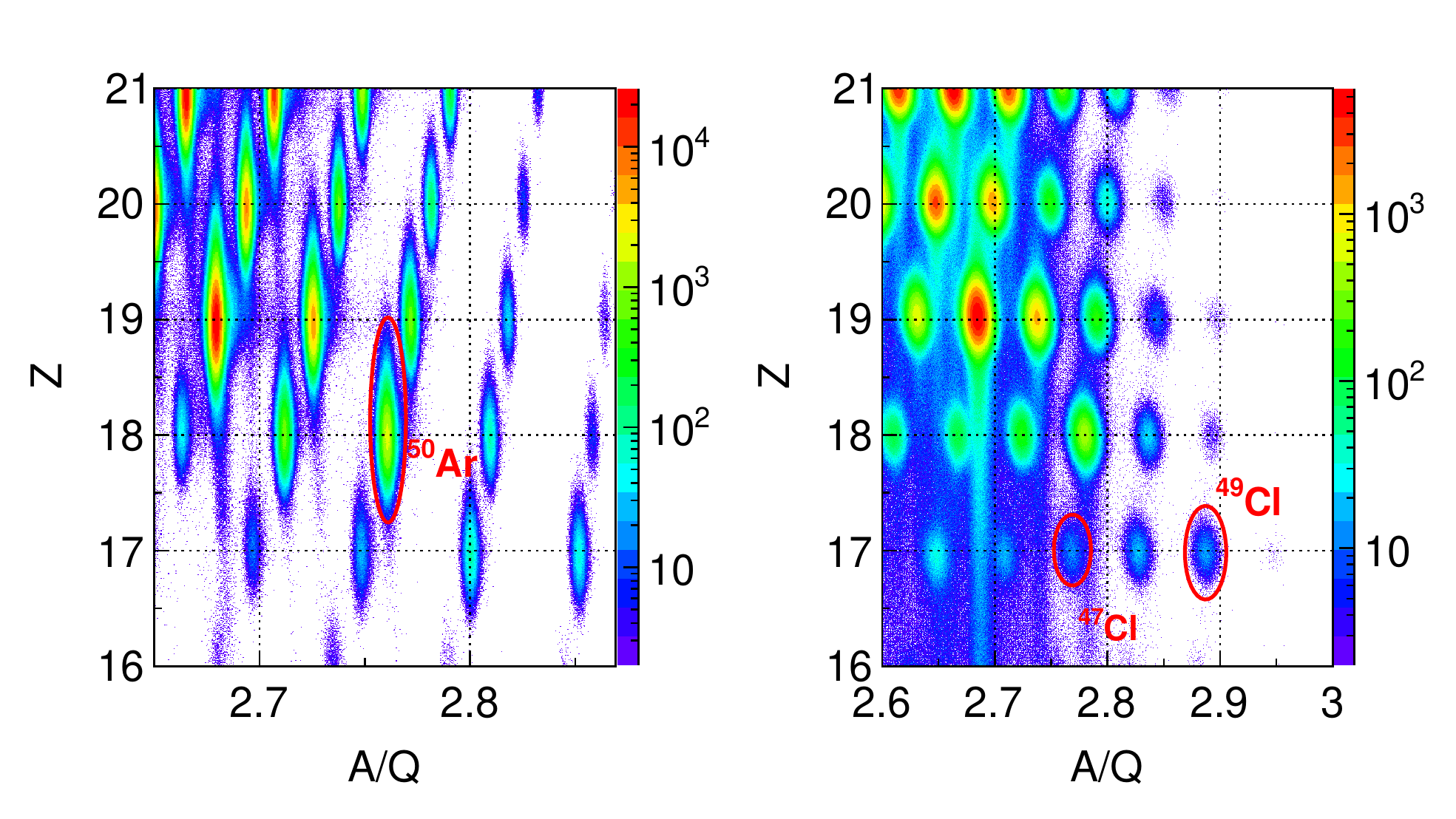}
		\end{center}
		\caption{Particle identification with the mass over charge number ratio A/Q and atomic number Z; (left) secondary beam particle identification at BigRIPS. The isotope $^{50}$Ar is indicated by the red ellipse. (right) Residue particle identification downstream the secondary target from the large acceptance SAMURAI spectrometer. The isotopes $^{47}$Cl and $^{49}$Cl are indicated by the red ellipses.} 
		\label{fig_ID}
	\end{figure}
	
	\section{Experimental setup}
	The experiment was performed at the Radioactive Isotope Beam Factory (RIBF), operated jointly by the RIKEN Nishina Center and the Center for Nuclear Study of the University of Tokyo. A $^{70}$Zn beam was accelerated up to 345 MeV/nucleon and impinged on a 10~mm-thick $^9$Be primary target at the entrance of the BigRIPS separator~\cite{kubo} with an average intensity of 240~pnA. The secondary beam was identified with magnetic rigidity $B\rho$, energy loss $\Delta E$ and time of flight $TOF$ measurements. The setting of the experiment was optimized for the study of the one-proton knockout reaction $^{53}$K(p,2p)$^{52}$Ar. Within the MINOS setup~\cite{ober}, a 151(1)~mm-thick liquid hydrogen target (LH$_2$) was used to compensate for the low intensity beams. The beam energy at the entrance (exit) of the secondary target was $\sim$247 ($\sim$184) MeV/nucleon, with an intensity of 2.9 particles/s for $^{50}$Ar. The total beam intensity on the target was about 212 particles/s. Scattered ions were analyzed with the SAMURAI spectrometer~\cite{koba} and identified by the mass over charge number ratio A/Q and atomic number Z on an event-by-event basis with the $B\rho - \Delta E - TOF$ method~\cite{fuk13}. Due to the large acceptance of the SAMURAI spectrometer, it was possible to measure the residues of many reaction channels in the same setup. Achieved resolutions before (after) the target were 0.057 \% (0.247) for $A/Q$ and 0.865 \% (0.726) for $Z$ with unambiguous separation of the different projectiles and residues, as shown in Fig.~\ref{fig_ID}.
	
	Prompt photons emitted at the MINOS target were detected with the DALI2\ts{+} array~\cite{take} composed of 226 NaI(Tl) detectors in a compact geometry. In order to optimize the energy resolution after Doppler correction, the vertex of the reaction in the target was determined with a cylindrical Time Projection Chamber (TPC) surrounding the target. Details of the MINOS setup are given in Ref.~\cite{ober}. 
	
	\section{Data analysis}
	\subsection{Determination of \texorpdfstring{$\gamma$}{Lg} ray energies}

	Each NaI detector of the DALI2\ts{+} array was calibrated individually using $^{133}$Ba, $^{137}$Cs, $^{60}$Co, and $^{88}$Y sources with good linearity from 356 to 1836 keV and an overall uncertainty $\sigma$ = 4 keV. The full-energy efficiency and energy resolution with add-back were determined using the GEANT4 framework~\cite{agos, SUNF}. They were found to be 30\% and 11\% (FWHM) for 1 MeV $\gamma$-rays emitted by particles moving at $\beta$ = 0.6, respectively. The GEANT4 application was used to provide a response function for each transition.
	
	The energies of $\gamma$ rays emitted at the target position from the residues at velocities close to $v/c$ = 0.6 have been corrected for the Doppler effect. This correction included the angle of the $\gamma$ rays measured with the DALI2\ts{+} array, the velocity of the projectile, and the reconstruction of the reaction vertex. The reaction vertices were determined from the tracks registered in the TPC for the protons emitted in the reaction~\cite{ober} and the beam track determined by drift chambers~\cite{koba}. Typical values for the vertex resolution were $\delta z_{v} = 5~\text{mm}$~\cite{santa2}, which corresponds to a time of flight resolution $\delta \tau $ = 30~ps. The reaction vertex may be different from the decay vertex if the lifetime of the populated state exceeds a few picoseconds. As mentioned in Ref.~\cite{santa2}, the width and shape of the measured photopeaks are sensitive to lifetimes above a few tens of picoseconds. 
	
	\subsection{\texorpdfstring{$\gamma$}{Lg} ray spectrum analysis}

	In previous similar analyses~\cite{santa,mur,liu3}, the background was fitted by two exponentials to take into account: $i)$ at high energy the unresolved background from partially detected high-energy transitions; $ii)$ at very low energy bremsstrahlung components due to electron-ion collisions. However, this procedure is not accurate enough when low energy transitions are involved, especially below 150~keV~\cite{npaul}. 
	
\begin{figure}[hbtp!]
		\begin{center}
			\includegraphics[width=8.6 cm]{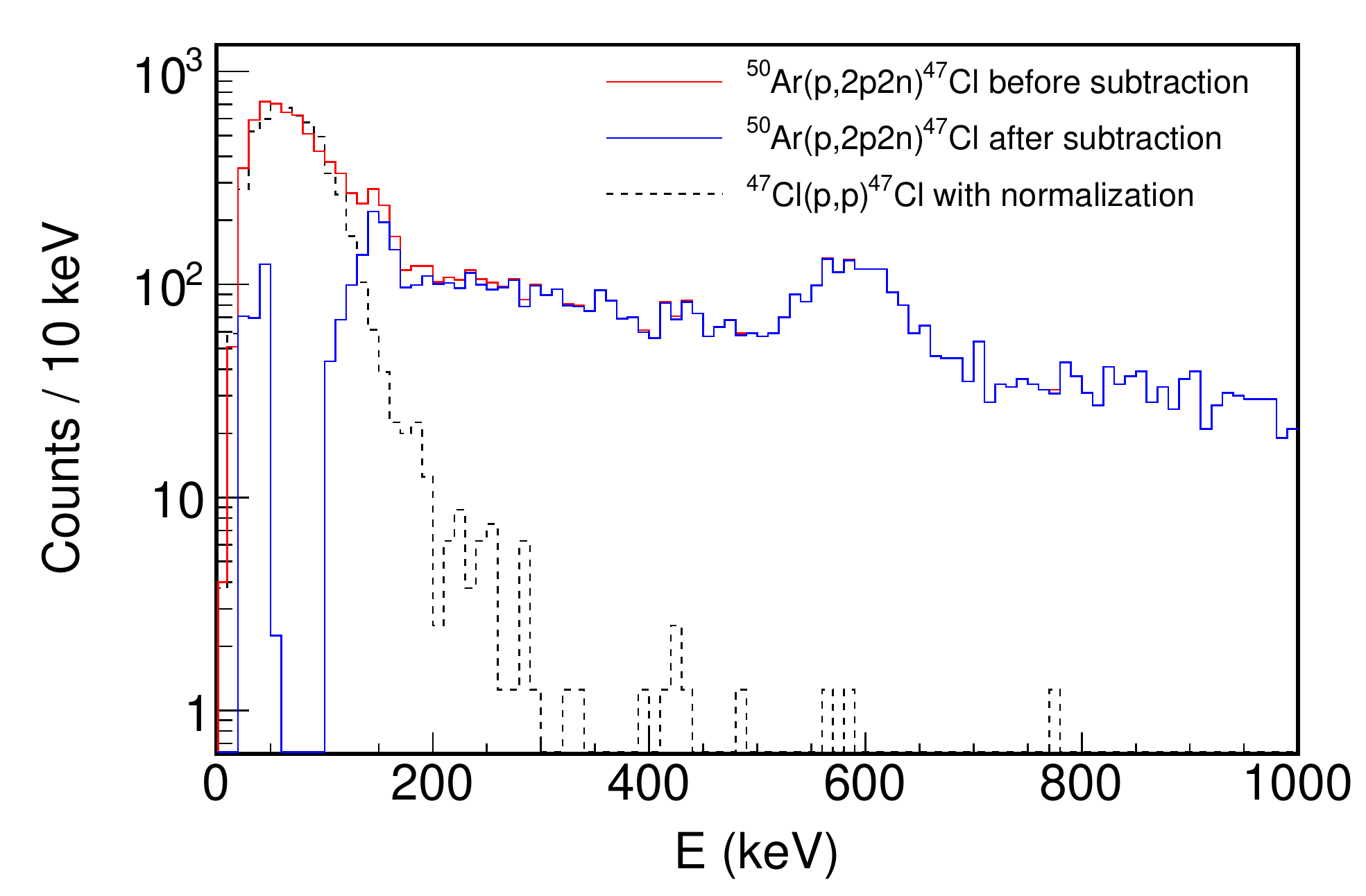}
		\end{center}
		\caption{Low-energy bremsstrahlung component subtraction: the red histogram is the total spectrum for $^{50}$Ar(p,2p2n)$^{47}$Cl; the blue histogram is the final spectrum after subtraction of the normalized black histogram for $^{47}$Cl(p,p)$^{47}$Cl.} 
		\label{fig2}
	\end{figure}

	In practice, the low energy bremsstrahlung component is obtained as the spectrum corresponding to the unreacted chlorine beam, namely the $^{A}$Cl(p,p)$^{A}$Cl channel. This component is subsequently normalized to the reaction of interest between 30 and 200 keV. The spectra for $^{50}$Ar(p,2p2n)$^{47}$Cl before and after subtraction, and the background, are shown in Fig. \ref{fig2} in the case of $^{47}$Cl for which the transition with the lowest energy is observed in this work at 148 keV. The same procedure is used for the knockout $^{50}$Ar(p,2p)$^{49}$Cl and subtraction from $^{49}$Cl(p,p)$^{49}$Cl.
	In the spectrum corresponding to the $^{A}$Cl(p,p)$^{A}$Cl channel, there is also a contribution from the inelastic excitation of the beam, which is about two orders of magnitudes smaller compared to atomic background (few tenths of mbarn vs. few barn) and can thus be neglected.
	
	Finally, the Doppler corrected $\gamma$-ray energy spectra are fitted by a combination of response functions and the above-determined background, as can be seen in Fig.~\ref{fig3}. For each transition, the final uncertainty on the centroid energy is obtained from the width of a $\chi^{2}$ distribution. Therefore, this width also includes the other uncertainty sources, such as errors in energy calibration with $\gamma$ sources and statistics. Lifetime effects are also included in the response functions. Increasing lifetimes result in decreasing the centroid energy and increasing the width of the full-energy peaks, as shown in Fig. 3.24 of Ref.~\cite{nanthe}.
	
	\subsection{Momentum distributions}
	
	Due to the large acceptance of the SAMURAI spectrometer, all the reaction products, and the unreacted beam were measured in the position-sensitive detectors used for identification, B$\rho$ reconstruction, and inclusive momentum distributions. For each momentum bin, the gamma-ray energy spectrum is obtained and fitted with response functions and background. This results in the parallel momentum distributions (PMD) and transverse momentum distributions (TMD) that may be seen in Fig.~\ref{fig4}. This procedure could be successfully applied only to the most intense transitions. Distributions for the ground state may be obtained from the difference between inclusive distributions and the sum of contributions from the most intense transitions, assuming negligible contributions from unresolved higher-lying states. This is particularly true for isotopes with small one-neutron separation energies, such as $^{49}$Cl where $S_{1n}$ = 3050(640) keV~\cite{ame}. 
	
	Both parallel and transverse momentum distributions obtained in this manner are sensitive to the angular momentum $\ell$ of the knocked-out nucleon and can be compared to distributions obtained from various reaction models. Besides the popular choice of distorted-wave impulse-approximation (DWIA)~\cite{bertu,wasa}, in the following, we also used the results of the transfer to continuum method (TC) with prior-form transition amplitudes in which the final state  is approximated by a continuum-discretized coupled-channels expansion of p-N states (p-p states for (p,2p) and p-n states for (p,pn)), as explained in detail in Ref.~\cite{moro}.
	
	\subsection{Cross sections}
	
	Inclusive cross sections were calculated from the number $N_{in}$ of projectiles entering the target and the number of ejectiles $N_{SAM}$ identified in the focal plane of the SAMURAI spectrometer as:
	\begin{equation}
		\sigma^{inc} (mb) = N_{SAM} / ( N_{in} * N_{T} * T)
		\label{sigmainc}
	\end{equation}
	
	\noindent
	with an overall transmission $T = 0.491(4)$, including the efficiencies of the beam detectors, the absorption of flux in the thick target, and the acceptance of the spectrometer. This value is obtained for each reaction as the ratio of identified outgoing residues versus projectiles for all trajectories which are well inside the spectrometer acceptance. The target density N$_{T}$ $ (cm^{-2})$ is given by
	\begin{equation}
		N_{T} = \rho * L * N_\mathrm{A}/m_H
	\end{equation}
	\noindent
	with the volumetric mass of liquid hydrogen $\rho$ = 70.973 g.cm$^{-3}$ at atmospheric vapor pressure, length $L$ of the target 15.15(10) cm, Avogadro number N$_\mathrm{A}$ and hydrogen mass m$_H$. Variations of the target density were controlled by an overall measurement of the vapor pressure. Charge state changes are not observed in the ionization chambers and therefore are not considered in this calculation, in agreement with the LISE++ calculations~\cite{tara}, predicting 1\% or less of charge states in this mass region.
	
	When statistics were high enough, exclusive cross sections were determined for each identified transition  $i \rightarrow f$, with the number $N^{\gamma_{0}}$ of photons taken as
	\begin{equation}
		N^{\gamma_{0}}_{i \rightarrow f} = \alpha_{i} / N_{in} 
	\end{equation}
	\noindent
 
	where the normalizing factor $\alpha_{i}$ was obtained from the fit with the DALI2\ts{+} response functions. 
	
	The exclusive cross section $\sigma^{ex}_{i}$ (mb) for the state $i$ is obtained as 
	\begin{equation}
		\sigma^{ex}_{i} = \sum_{f} \: N^{\gamma_0 corr}_{i \rightarrow f} / ( N_{in} * N_{T} * \varepsilon_{MINOS} * T)
		\label{sigmaex}
	\end{equation}
	
	\noindent
	where $\varepsilon_{MINOS}$ is the efficiency of detecting at least one proton in the TPC obtained in a simulation with a 15 cm long target. $\varepsilon_{MINOS}$ = 0.89(2) was found for the $^{50}$Ar(p,2p) reaction. An effective $N^{\gamma_0 corr}_{i \rightarrow f}$ was used, subtracting $N^{\gamma_{1..N}}_{i \rightarrow f}$ from $N^{\gamma_0}_{i \rightarrow f}$ to take into account the feeding from higher-lying states 1...N, when they can be identified in the energy spectrum and the level scheme. Since the feeding from non-identified transitions cannot be accounted for, the obtained values are upper limits of exclusive cross-sections for one proton knockout to the given state.
	
	\section{Spectroscopy of \texorpdfstring{$^{49}$C\lowercase{l}}{Lg} from the one-proton knockout channel \texorpdfstring{$^{50}$A\lowercase{r(p,2p)}}{Lg}}
	
	In the simplest shell-model framework, ten protons in $^{50}$Ar occupy the $sd$ shell valence space with three active orbitals, ${\pi}0d_{5/2}, {\pi}1s_{1/2}$, and ${\pi}0d_{3/2}$. Then, one-proton knockout reactions exclusively populate positive parity states in $^{49}$Cl, which have a sizable overlap with proton-hole configurations ${\pi}(1s_{1/2})^{-1}$ and ${\pi}(0d_{3/2})^{-1}$ in $^{50}$Ar. The relative position of the two valence orbitals is sensitive to the details of the proton-neutron interaction with a possible inversion of the ground state spin ${1/2}^{+}$ versus ${3/2}^{+}$, as already observed for $^{47,49,51}$K~\cite{papu1,papu2}.
	
	The collective ${2}^{+}_{1}$ state was found at 1150(12) keV in $^{50}$Ar~\cite{cor}, consistent with~\cite{step2}. States resulting from the coupling of the ${2}^{+}$ to the ${3/2}^{+}$ or ${1/2}^{+}$ states may be present in the level scheme of $^{49}$Cl around this energy but are expected to be weakly populated in the direct one-proton knockout reaction.
	
	\subsection{ Experimental results}
	
\begin{figure}[hbtp!]
		\includegraphics[ width=8.6cm]{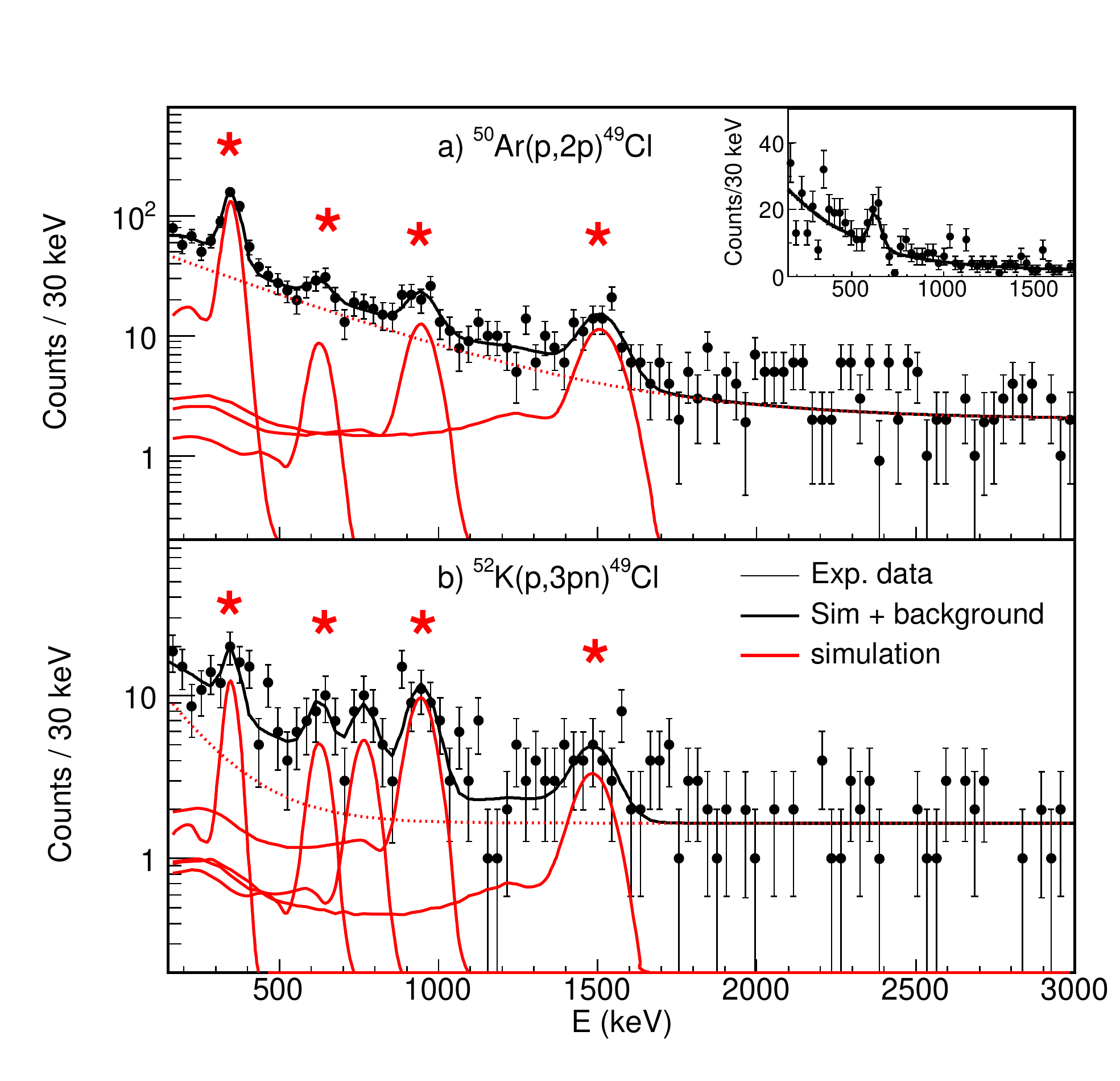}
		\caption{a) Doppler corrected $\gamma$-ray energy spectrum populated from $^{50}$Ar(p,2p)$^{49}$Cl for all $\gamma$ multiplicities after subtraction of the low energy bremsstrahlung component. Experimental data (points) are fitted by a combination (black line) of four DALI2\ts{+} simulated response functions (red continuous lines) and a two-component exponential background (red dashed line). Four transitions are identified by stars at 350, 630, 970 and 1515 keV. The inset shows the spectrum gated on the 350~keV transition after subtraction of a component gated at higher energy; 
			b) the same spectrum for the non-direct reaction channel $^{52}$K(p,3pn)$^{49}$Cl, analyzed with response functions at the same  energies.
		}
		\label{fig3} 
	\end{figure}

	Bound states may be observed by $\gamma$-ray emission up to the one-neutron separation energy $S_{1n}$ = 3050 (640) keV~\cite{ame} for $^{49}$Cl. Doppler corrected $\gamma$-ray energy spectra for $^{49}$Cl are shown in Fig.~\ref{fig3} after subtraction of the low energy bremsstrahlung component. The one-proton knockout channel $^{50}$Ar(p,2p), which favors single particle states, is compared to a more complex reaction channel $^{52}$K(p,3pn), for which a different population of states with a stronger np-nh component is expected. 
	
	In the $^{50}$Ar(p,2p) reaction, a strong transition is observed at 350 keV, while weaker structures may be observed at 630, 970, and 1515 keV. There is no strong evidence for any transition at higher energy above the exponential background below $S_{1n}$. A confidence level analysis performed on the single gamma spectrum confirms the 350, 630, 970, and 1515 keV transitions with confidence values of 7., 3.0, 3.5, and 5.5 $\sigma$. 
	
	The black line in Fig.~\ref{fig3}-a) is the final fit to the (p,2p) channel with response functions corresponding to transitions at 350(6), 630(15), 970(27), and 1515(32) keV. Only prompt transitions were considered here. 
	
	For all $\gamma$-ray multiplicities, a $\gamma$-$\gamma$ analysis was performed with four gates corresponding to the main transitions and used to produce coincidence spectra. Due to low statistics, we only considered the spectrum gated by the range corresponding to the first transition at 350 keV, shown in the inset of Fig.~\ref{fig3}. The confidence level for a coincidence with the transition at 630 keV was 5$\sigma$. The summed energy of the coincidence is 980(16) keV, very close to the energy of the 970(27) keV transition observed in the singles spectrum. To investigate a possible direct decay to the ground state by a transition at 980 keV, the singles spectrum Fig.~\ref{fig3}-a) was also analyzed with two response functions at 970 and 980 keV, assuming equal weights which maximizes the broadening of the resulting structure.
	Within the energy resolution of the DALI2\ts{+} array, there is no significant difference between the two analyses of Fig.~\ref{fig3}-a). Therefore, it is not possible experimentally to discriminate between two separate states at 970(27) and 980(16) keV or only one state.
	We do not see evidence for a coincidence with the other strong transition present in the singles spectrum at 1515 keV, suggesting a direct ground state decay.
	
	These results can be compared to the spectrum obtained in the multi-nucleon removal $^{52}$K(p,3pn)$^{49}$Cl, for which a direct population of single particle states is not generally expected. The Doppler-corrected energy spectrum in Fig.~\ref{fig3}-b) has been analyzed with response functions at the same energies used for the (p,2p) reaction, except for a new weak transition at 768(22) keV added to improve the fit. Without further information, it is not possible to place the 768 keV transition in the level scheme shown in Fig.~\ref{fig5}. Due to the weak intensity of the 350 keV transition and limited statistics, the spectrum gated by a range around 350 keV is not conclusive in this case.
	\begin{center}
		\begin{table}[hbtp!]
			\caption{
Transitions observed in Fig.~\ref{fig3} for the two differents reaction channels $^{50}$Ar(p,2p)$^{49}$Cl and $^{52}$K(p,3pn)$^{49}$Cl: excitation energy $E^{*}$, detection-efficiency corrected intensity $I_{i}$ and ratio $I_{i}/I_{970}$ normalized to the 970 keV transition.}
			\begin{ruledtabular}
			\begin{tabular*}{0.9\textwidth}{@{\extracolsep{\fill}}ccccccccccc}
				& \multicolumn{2}{c}{$^{50}$Ar(p,2p)$^{49}$Cl}& \multicolumn{2}{c}{$^{52}$K(p,3pn)$^{49}$Cl}\\
				\cmidrule(lr){2-3}\cmidrule(lr){4-5} 
				$E^{*}$ &$I_{i}$&$I_{i}/I_{970}$&$I_{i}$&$I_{i}/I_{970}$&\\
				(keV)&\\
				\midrule
				350 (6)&355(19)&5.55(30)& 24(5)&0.67(14)\\
				630 (15) &36(6)&0.56(9)&14(4)&0.39(11)\\
				768 (22)&&&16(4)&0.44(11)\\
				970 (27)&64(8)&1.00(12)&36(6)&1.00(16)\\
				1515 (32)&67(8)&1.05(12)&18(4)&0.50(11)\\
			\end{tabular*}
			\end{ruledtabular}
			\label{tab1}
		\end{table}
	\end{center}
	\begin{center}
		\begin{table*}[htbp!]
			\caption{Spin, excitation energies $E^{*}$, spectroscopic factors $C^{2}S$ and cross sections $\sigma_{th,i}$ for $^{49}$Cl states: theoretical values for the one-proton knockout reaction $^{50}$Ar(p,2p) are given for the shell-model calculations with the SDPF-MU and SDPF-MU$_s$ interactions, and IMSRG calculation with the 1.8/2.0 (EM) interaction (see Section VI). Theoretical cross sections $\sigma_{th,i}$ are obtained from Eq.~(\ref{sigma}) using $C^{2}S$ and single-particle cross sections $\sigma_{sp}^{lj}(E^{*},E_{inc})$ from the TC and DWIA methods described in text and given in the two last columns. The excitation energy range is limited to 2000 keV, as no state above this energy with a sizeable spectroscopic factor has been obtained in the theoretical calculations. A correspondence with experimental data is also proposed. The last row is the sum of exclusive cross sections $\sum \sigma_{i}^{ex}$ for the three employed calculations to be compared to the measured inclusive cross section $\sigma_{inc}$.}
			\begin{ruledtabular}
			\begin{tabular*}{0.9\textwidth}{@{\extracolsep{\fill}}cccccccccccccccccc}
				&\multicolumn{4}{c}{SDPF-MU} &\multicolumn{3}{c}{SDPF-MU$s$}
				&\multicolumn{3}{c}{IMSRG(1.8/2.0 EM)}
				&\multicolumn{3}{c}{Experiment}
				&\multicolumn{3}{c}{$\sigma_{sp}^{lj}(E^{*},E_{inc})$}\\
				\cmidrule(lr){2-5}\cmidrule(lr){6-8}\cmidrule(lr){9-11}\cmidrule(lr){12-14}\cmidrule(lr){15-17} 
				State&$E^{*}$&$C^{2}S$&$\sigma_{th,TC}$&$\sigma_{th,DW}$&$E^{*}$&$C^{2}S$&$\sigma_{th,TC}$&$E^{*}$&$C^{2}S$&$\sigma_{th,TC}$
				&$E_{exp}$&$\sigma_{exp}$&$\Delta\ell$&$nlj$&$\sigma_{TC}$&$\sigma_{DW}$\\
				&(keV)&&(mb)&(mb)&(keV)&&(mb)&(keV)&&(mb)&(keV)&(mb)&&(mb)&(mb)\\
				\midrule
				3/2$^{+}_{1}$&  gs&1.910&3.04&2.71&  83&2.026&3.22&  gs&2.527&4.02
				&gs& $<$2.26(18)&2&$0d3/2$&1.59&1.42 \\
				1/2$^{+}_{1}$&   419&1.396&2.18&2.25&gs&1.272&1.98& 135&1.055&1.65&
				350(6)&1.25(9)&0&$1s1/2$&1.56&1.61 \\
				3/2$^{+}_{2}$&   1454&0.206&0.33&0.29&849&0.184&0.29& 724&0.015&0.02&&\\
				5/2$^{+}_{1}$&   1248&0.030&0.05&0.05&922&0.040&0.07& 991&0.002&0.&&\\
				7/2$^{+}_{1}$&1745&&&&1727&&&1477&&&&\\
				5/2$^{+}_{2}$&1701&0.515&0.85&0.80&1660&0.422&0.70&1762&0.453&0.75&1515(32)&0.55(4)&&$0d5/2$&1.66&1.56\\
				\midrule
				$\sum \sigma_{i}^{ex}$& & &  6.45 & 6.1 & & & 6.26 & & & 6.44\\
				&&&&&&&&&&&&$\sigma_{incl}$ = 4.55(15)\\
			\end{tabular*}
			\end{ruledtabular}
			\label{tab2}
		\end{table*}
	\end{center}
	\begingroup
	\renewcommand*\arraystretch{1.2}
	\begin{center}
		\begin{table*}[ht]
			\caption{Energies, B(E2$\downarrow$) and B(M1$\downarrow$) values for transitions between low-lying states obtained in the SDFP-MU shell-model calculation for $^{49}$Cl.}
			\begin{tabular*}{0.6\textwidth}{@{\extracolsep{\fill}}cccccc}
				\toprule
				\toprule
				\multirow{2}{*}{$^{49}$Cl}&Energy& $B(E2)$$\downarrow$&$\tau$ & B(M1)$\downarrow$&$\tau$\\
				& (keV)&e$^{2}$fm$^{4}$&(ps)&$\mu_{N}^{2}$&(ps)\\
				\midrule
				1/2$^{+}_{1}$  $\rightarrow$  3/2$^{+}_{1}$&419&10.6 &$>$ 1ns& 0.021&37.7\\
				\midrule
				5/2$^{+}_{1}$  $\rightarrow$  3/2$^{+}_{1}$&1248&93.8&2.9&0.001&24.4\\
				
				5/2$^{+}_{1}$  $\rightarrow$  1/2$^{+}_{1}$&829&108.9&19.2&&\\
				\midrule
				3/2$^{+}_{2}$  $\rightarrow$  3/2$^{+}_{1}$&1454&8.6&14.7&0.005&3.7\\
				
				3/2$^{+}_{2}$  $\rightarrow$  1/2$^{+}_{1}$&1035&124.0&5.6&0.028&1.8\\
				
				3/2$^{+}_{2}$  $\rightarrow$  5/2$^{+}_{1}$&206&24.7&$>$ 1 ns&0.129&50.2\\
				\midrule
				5/2$^{+}_{2}$  $\rightarrow$  3/2$^{+}_{1}$&1701&11.7&4.9&0.089&0.1\\
				
				5/2$^{+}_{2}$  $\rightarrow$  1/2$^{+}_{1}$&1282&67.8&3.5&&\\
				
				5/2$^{+}_{2}$  $\rightarrow$  5/2$^{+}_{1}$&453&4.3&$>$ 1 ns&0.021&29.1\\
				
				5/2$^{+}_{2}$  $\rightarrow$  3/2$^{+}_{2}$&247&2.4&$>$ 1 ns&0.029&132.3\\
				\bottomrule
				\bottomrule
			\end{tabular*}
			\label{tab8}
		\end{table*}
	\end{center}
	\endgroup
	\vspace{-1.5 cm}
	
	Based on the small spectroscopic factors reported in Tab.~\ref{tab2}, the 970 keV state could be the $3/2^{+}_{2}$ or $5/2^{+}_{1}$ state populated in a non-direct process and decaying to the ground state. If we take this state as a reference, the detection-efficiency corrected intensity ratio $I_{i}/I_{970}$ displayed in Tab.~\ref{tab1} for the two reactions $^{50}$Ar(p,2p)$^{49}$Cl and $^{52}$K(p,3pn)$^{49}$Cl provide information on the direct population of a state decaying by the transition $i$. The ratio $I_{350}/I_{970}$ is 5.5(7) and 0.7(2), respectively, and shows the single particle character of the 350 keV state. The ratio $I_{1515}/I_{970}$ is 1.0(2) and 0.5(2), leading to a similar but weaker conclusion for the 1515 keV state. Finally, the $I_{630}/I_{970}$ ratio is 0.39(12) and 0.56(12), corresponding to a similar ratio for the two reactions populating the 970 keV state.
	
	According to the level scheme proposed in Fig.~\ref{fig5}, we determined the experimental cross sections $\sigma_{exp}$ displayed in Tab.~\ref{tab2}.	For each transition, an exclusive cross section can be calculated from Eq.~(\ref{sigmaex}), yielding values of 1.25(9), 0.49(4), and 0.55(4) mb for the 350, 970, and 1515 keV states, respectively. The inclusive cross section for the $^{50}$Ar(p,2p)$^{49}$Cl reaction was determined with Eq.~(\ref{sigmainc}) to be $\sigma_{inc}$ = 4.55(15) mb. Then, the cross section $\sigma_{gs}$ to populate the ground state was deduced as the difference between the inclusive cross section $\sigma_{incl}$ and the sum of the excited-state cross sections $\Sigma \: \sigma^{ex}_{i}$~=~2.29(10) mb for transitions known to feed the ground state, so that $\sigma_{gs}$ = 2.26(18) mb. It is only an upper limit for $\sigma_{gs}$, assuming that the feeding from unresolved higher lying states can be neglected, which is reasonable considering the low value of $S_{1n}$. 
	
	Experimental values are compared in Tab.~\ref{tab2} to the results of cross section calculations $\sigma^{ex}_{i}(E^{*})$ for excitation energies $E^{*}$ at a given incident energy $E_{inc}$ following
	\begin{equation}
		\sigma^{ex}_{i}(E^{*}) = \sum\limits_{l,j} C^{2}S^{i}_{l,j} * \sigma_{sp}^{lj}(E^{*},E_{inc})
		\label{sigma}
	\end{equation}
	as the product of theoretical spectroscopic factors $C^{2}S^{i}_{l,j}$ with single particle cross sections $\sigma_{sp}^{lj}(E^{*},E_{inc})$ obtained in a reaction model. $\sigma_{sp}^{lj}(E^{*},E_{inc})$ values were calculated with the TC~\cite{moro} and DWIA~\cite{wasa} methods for the $^{50}$Ar(p,2p) reaction at 217 MeV/nucleon, which corresponds to the mid-target energy for $^{50}$Ar. Due to the large target thickness, there was a significant variation of projectile energy from 247 MeV/nucleon at the entrance down to 184 MeV/nucleon at the exit. The variation of the cross section through the target was carefully accounted for in the similar study of~\cite{liu}, but the final values were not found to be different from mid-target values by more than 1\%. This is in agreement with the mean value $<\sigma>$ calculated for the $0d_{3/2}$ orbital from the entrance to exit and found to be 1.2\% smaller than the mid-target value, which has thus a very limited impact on our comparison with theoretical values.
	Single-particle $\sigma_{sp}^{lj}(E^{*},E_{inc})$ were calculated with the TC and DWIA methods for the removal of a proton in the different orbitals and energies ($0d_{3/2}$, 0 keV), ($1s_{1/2}$, 350 keV), and ($0d_{5/2}$, 1515 keV). Values are given in Tab.~\ref{tab2} and used in the calculation of $\sigma_{th}$ for the $3/2^{+}_{1}$, $1/2^{+}_{1}$, and $5/2^{+}_{2}$ states. Due to the weak dependence with $E^{*}$, the same values were used for $3/2^{+}_{2}$ and $5/2^{+}_{1}$. These two methods were recently benchmarked with the one neutron knockout reaction $^{15}$C(p,pn) at 420~MeV/nucleon, and differences in cross section were found to be below 5\%~\cite{yos}.

	Overall consistency may be tested through the reduction factor $R_{s}$ = $\sigma_{inclusive}$ /$\sum \sigma_{i}^{ex}$, using the spectroscopic factor predictions of a shell model routinely used in this region like SDPF-MU in Tab.~\ref{tab2}. The values $R_{s}$ = 0.70(2) and 0.75(2) are found with the TC and DWIA reaction models, respectively, which places $^{50}$Ar(p,2p) at $\Delta S$ = 17.0~MeV in the general trend observed for the one nucleon knockout reactions~(see Fig.2 in Ref.~\cite{gom}).
	
	This correspondence between experimental data and both TC and DWIA calculations justifies the underlying single particle character of the populated states. However, it is not possible, at this step, to identify the spin parity of the ground state and first excited state.
	
\begin{figure*}
		\begin{center}
			\includegraphics[width=12.6cm]{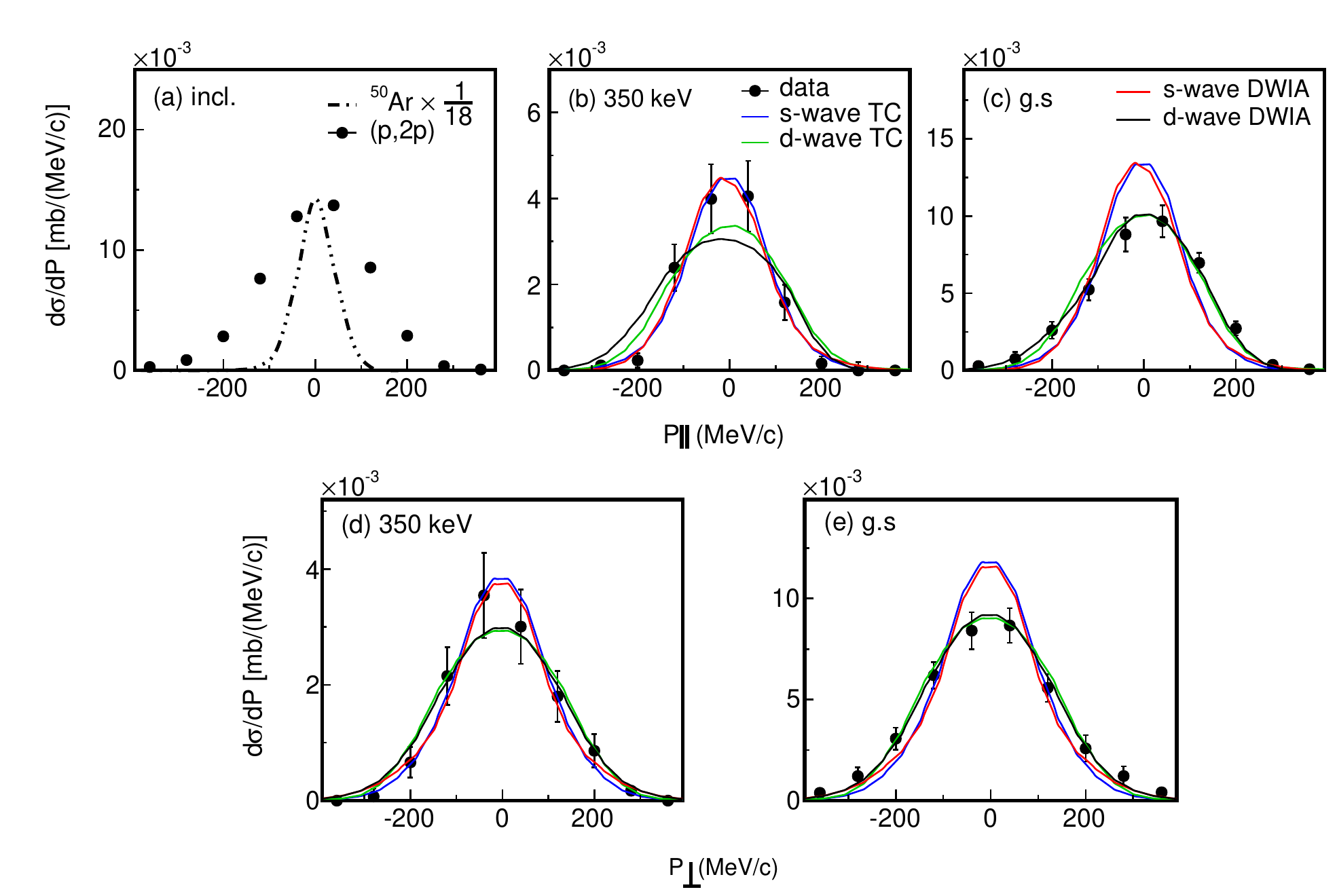}
		\end{center}
		\caption{Momentum distributions of $^{49}$Cl ejectiles measured with the SAMURAI spectrometer: a) PMD of unreacted beam and inclusive $^{50}$Ar(p,2p)$^{49}$Cl reaction; PMD b) in coincidence with the 350 keV transition measured in DALI2\ts{+}; c) determined for the ground state; TMD d) in coincidence with the 350 keV transition measured in DALI2\ts{+}; e) determined for the ground state. Data are compared to calculations with the TC and DWIA methods for $\ell =0$ and $\ell =2$ waves after convolution with the experimental resolution.
	}
		\label{fig4} 
	\end{figure*}

\begin{figure}[hbtp!]
		\includegraphics[width=8.6cm]{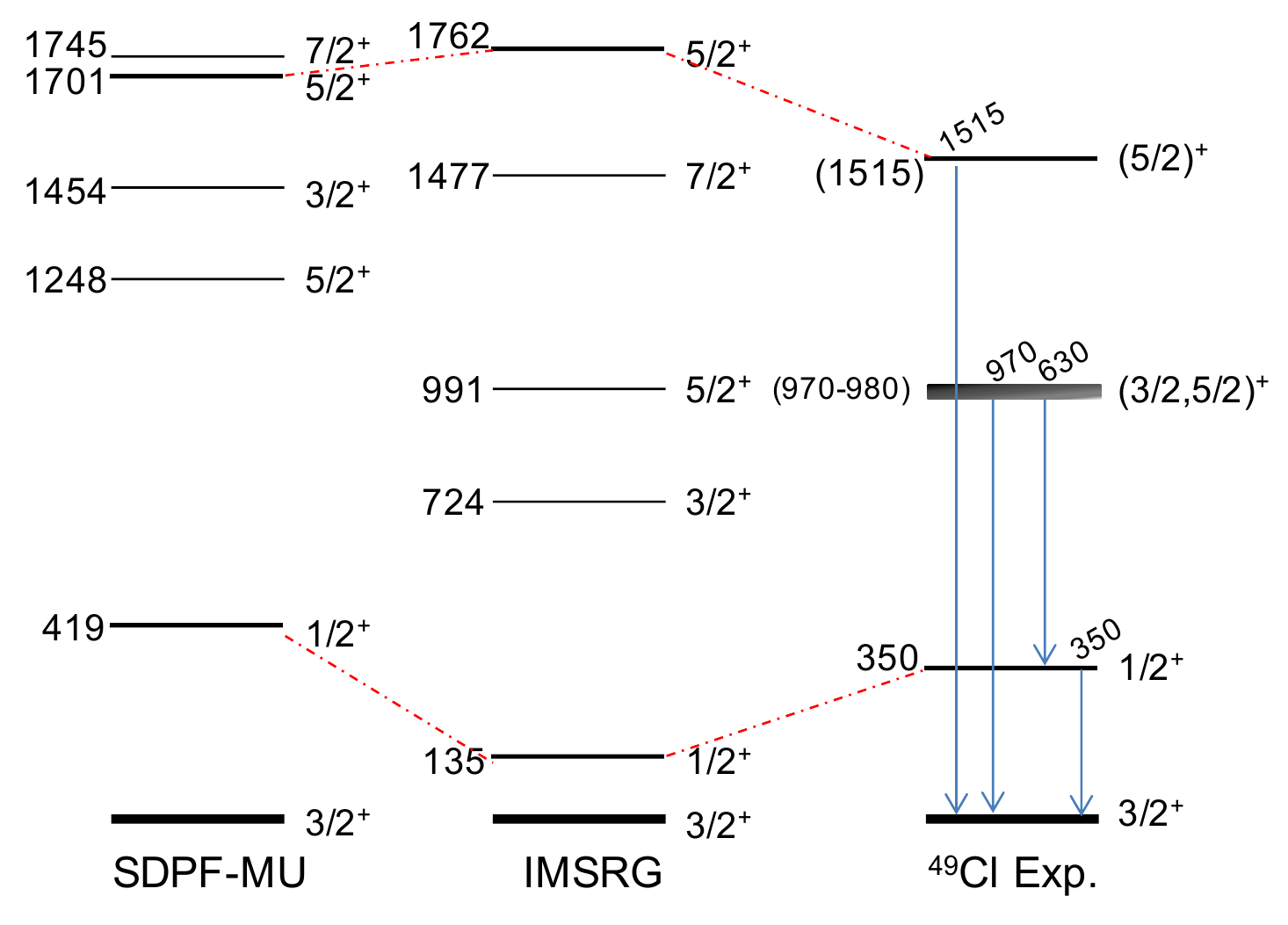}
		\caption{Level scheme of $^{49}$Cl compared to theoretical calculations detailed in section VI. The shaded area stands for the possible existence of a single or two different states (see text). 
		}
		\label{fig5} 
	\end{figure}

	The transverse and parallel momentum distributions of the ejectiles could be obtained with the SAMURAI spectrometer and its associated detectors, as shown in Fig.~\ref{fig4}. PMD for the unreacted beam and inclusive (p,2p) reaction are shown in Fig.~\ref{fig4}-a). Due to limited statistics for exclusive measurements, TMD and PMD could only be extracted for the ground state and first excited state displayed in Fig.~\ref{fig4}-b)-d). For each momentum value, the amplitude of the 350 keV transition was determined from the gamma-ray fit procedure (see Fig.~\ref{fig3}). Inclusive data were used after appropriate subtraction to obtain the distribution for the ground state.

	The TC and DWIA methods were used to calculate the PMD and TMD for the one proton knockout $^{50}$Ar(p,2p) reaction in both cases, with $0d_{3/2}$ $\ell =2$ and $1s_{1/2}$ $\ell =0$. The width of the unreacted beam ($\sigma$ = 38 MeV/c) is used for the convolution of the theoretical TMD and PMD distributions. After convolution, TMD and PMD are compared to data in Fig.~\ref{fig4}-c)-e), providing evidence for the $\ell =2$ character of the ground state. Fig.~\ref{fig4}-b)-d)) instead suggests a $\ell = 0$ character for the narrower distribution associated with the 350 keV transition. A further test was done with a Bayesian analysis~\cite{rob1995}. We find that the log$_{10}$ scaled Bayes factors is $log_{10}(B_{10})$ > 7 \footnote{Bayes factors provide decisive evidence for one model when compared to another model if their log$_{10}(B_{10})$ is larger than 2.} such that a d-wave character is preferred over s-wave in the ground state distributions, and s-wave over d-wave in the 350-keV state distributions, which quantitatively supports our $\ell$ hypothesis.
	
	\subsection{Level scheme of \texorpdfstring{$^{49}$Cl}{Lg}}
	
	Based on these observations, a level scheme for $^{49}$Cl is proposed in Fig.~\ref{fig5}. A spin parity $J^{\pi} = 3/2^{+}$ is assigned for the ground state, based on the momentum distribution and partial level cross section. The first excited state decays to the ground state by the transition at 350 keV; a spin-parity $J^{\pi} = 1/2^{+}$ is assigned from the momentum distribution and the partial level cross section, which is lower than the ground state cross section in all theoretical calculations. A state, possibly $J^{\pi} = (3/2,5/2)^{+}$, is proposed at 970(27)~keV which directly decays to the ground state. From $\gamma-\gamma$ coincidences, another state is suggested at 980(16) keV, decaying to the first excited state. These two energies are very close to each other and compatible with only one state. Since we cannot firmly rule out the existence of two different states, a shaded area in Fig.~\ref{fig5} stands for the unresolved existence of two different states or a single state. At similar energies, the SDPF-MU and IMSRG calculations also predict the existence of two possible states, $3/2^{+}_{2}$ and $5/2^{+}_{1}$, with small spectroscopic factors (Tab.~\ref{tab2}), suggesting a weak direct population in the one-proton knockout reaction. In the SDPF-MU calculation, the $5/2^{+}_{1}$ decays mainly tho the ground state, as shown in Tab.~\ref{tab8}. Finally, a spin-parity of $J^{\pi} = 5/2^{+}$ is proposed here for the level at 1515~keV and a direct decay to the ground state. This assignment is based on the spectroscopic factor obtained in the SDPF-MU and ISMRG calculations for a $5/2^{+}_{2}$ state near 1700 keV (Tab.~\ref{tab2}), corresponding to a sizable part of the total strength. This is consistent with the shell model calculations of ($BE2$) and ($BM1$) values, which predict a strong decay from $5/2^{+}_{2}$ to $3/2^{+}_{1}$ by M1 transition (Tab.~\ref{tab8}).

	\section{Spectroscopy of \texorpdfstring{$^{47}$C\lowercase{l}}{Lg}}
	
	In the same experiment, $^{47}$Cl reaction residues were also transmitted through the SAMURAI spectrometer. However, $^{48}$Ar was poorly transmitted through BigRIPS, resulting in few events for the one-proton knockout $^{48}$Ar(p,2p)$^{47}$Cl reaction. Therefore, neither momentum distributions nor spin assignment could be obtained for $^{47}$Cl. Other projectiles were better transmitted to the target, resulting in various reaction channels, either multi-nucleon removal reactions like $^{50}$Ar(p,2p2n)$^{47}$Cl or the one-neutron knockout reaction $^{48}$Cl(p,pn)$^{47}$Cl. In $^{47}$Cl, the one-neutron separation energy $S_{1n}$ =3920 (220) keV~\cite{ame}.
	
\begin{figure}[hbtp!]
		\includegraphics[width=8.6cm]{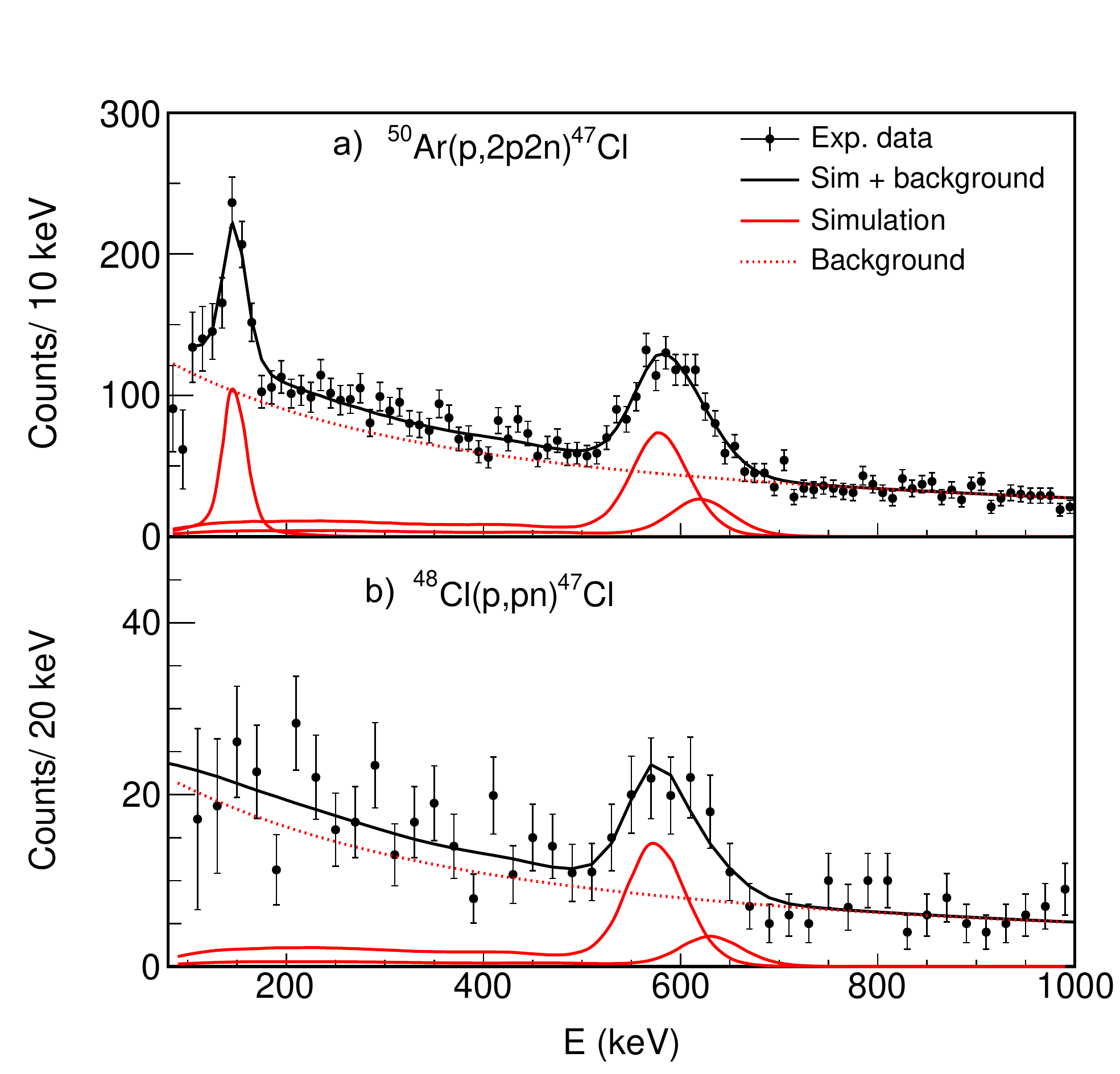}
		\caption{ Doppler corrected $\gamma$-ray energy spectrum of $^{47}$Cl produced from : a) the $^{50}$Ar(p,2p2n) multi-nucleon removal reaction. The broad structure around 600 keV is reproduced with similar $\chi^{2}$, either with two transitions at 578 keV and 634 keV and no lifetime, or a single transition at 624 keV and a lifetime $\tau$ = 130 ps; b) the neutron knockout reaction $^{48}$Cl(p,pn. Response functions at the same energies are used in the fits displayed in the top and bottom panels. }
		\label{fig7} 
	\end{figure}

	In Fig.~\ref{fig7} the bremsstrahlung-subtracted Doppler corrected $\gamma$-ray energy spectrum corresponding to the $^{50}$Ar(p,2p2n)$^{47}$Cl reaction is shown and fit with a double exponential background and simulated response functions. One can clearly observe: $i$) a sharp peak at 148(4) keV consistent with a negligible lifetime (less than 50 ps); $ii$) a broader structure around 600 keV. The analysis of the broad structure is not unique, and we find a good fit with an equivalent $\chi^{2}$ value either with $i$) two response functions at 578(12) and 634(23) keV and no lifetime, as shown in Fig.~\ref{fig7}; $ii$) a single peak at 624(7) keV with a lifetime around 130 ps. From a $\gamma-\gamma$ analysis, there is no evidence for a coincidence between the 148 keV transition and the broad structure. Therefore, we suggest a state at low excitation energy $E^{*}$ = 148 keV. A cascade between the two transitions at 578 and 634 keV can be ruled out since there is no evidence for a coincidence in the experimental data.
	
	\begin{figure}[hbtp!]
		\begin{center}
			\includegraphics[width=8.6cm]{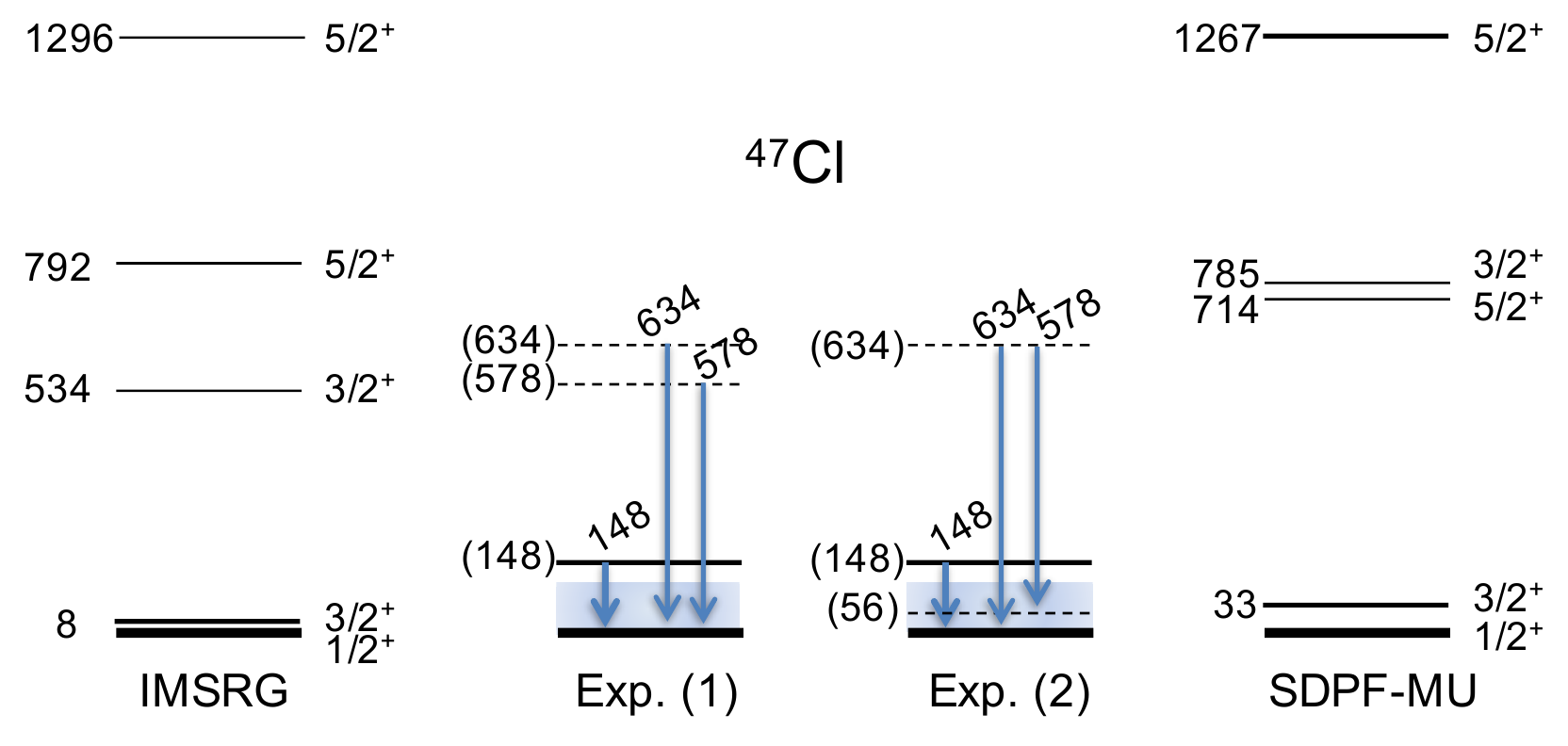}
		\end{center}
		\caption{Two different experimental level schemes proposed for $^{47}$Cl, compared to theoretical predictions. The shaded area below 100 keV was not accessible in the experiment.}
		\label{fig9}
	\end{figure}
	
	The $^{48}$Cl secondary beam was also transmitted to the LH$_{2}$ target. The Doppler corrected $\gamma$-ray energy spectrum from the one-neutron knockout reaction $^{48}$Cl(p,pn)$^{47}$Cl is shown in Fig.~\ref{fig7}-b). The same transition energies are used here as for the spectrum above. Due to low statistics, it is again impossible to conclude whether there are one or two peaks around 600 keV. The low-lying 148 keV transition is surprisingly not visible in the one-neutron knockout reaction, which is probably due to its selectivity compared to the multi-nucleon removal, i.e. the state populated in the multi-nucleon removal does not have a neutron-hole single-particle character.
	
	The inclusive cross section for the $^{48}$Cl(p,pn)$^{47}$Cl reaction was determined to be 19.3(15)~mb. Cross sections corresponding to the structure at excitation energy around 600 keV have been determined, depending on the one or two transition analyses: $i$) 10.0(13)~mb and 2.7(5)~mb for two states at 578 keV and 634 keV; $ii$) 11.4(15)~mb for only one state at 624 keV. For the direct population of the ground state, an upper limit was obtained by difference, with 6.6(20)~mb and 7.9(21)~mb for the two scenarios, respectively. No information could be obtained on spin assignments from the momentum distributions of $^{47}$Cl residues due to the complex non-direct multi-nucleon removal reaction $^{50}$Ar(p,2p2n), and the low statistics for the $^{48}$Cl(p,pn) reaction. 
	
	Due to the DALI2\ts{+} energy threshold, the decay of a state at very low excitation energy (below 100 keV, corresponding to the shaded area in Fig.~\ref{fig9}) could not be determined for this experiment. Considering all uncertainties, the level scheme proposed in Fig.~\ref{fig9} is restricted to a state at 148(4) keV without spin assignment. 
	We again mention the possibility for two states (dashed lines) at excitation energies of 578(12) and 634(23) keV, assuming 0 ps lifetime and without spin-parity assignment. Both the SDPF-MU and IMSRG calculations predict the existence of $3/2^{+}_{2}$ and $5/2^{+}_{1}$ states with excitation energies compatible with the measured transitions. 
	
	\section{Discussion}
	\label{discuss} 
	\subsection{Theoretical approaches}
	Data are compared to several state-of-the-art theoretical calculations. 
	
	The first one is a large scale shell-model calculation employing the SDPF-MU Hamiltonian~\cite{utsu}, performed using the KSHELL code detailed in Ref.~\cite{shi}. The valence space contains the $sd$ orbitals for protons and the $fp$ orbitals for neutrons, with a $^{28}$O inert core. The main interactions are USD~\cite{brown} for protons and the GXPF1B Hamiltonian~\cite{honma} for neutrons. The cross-shell part is given by $V_{MU}$ of Ref.~\cite{otsuka}. Of particular importance is the reproduction of spectroscopic factors of proton $sd$ orbitals for $^{48}$Ca, with the inclusion of a tensor force, as shown in Ref.~\cite{utsu}. A good agreement is obtained for neutron-rich calcium isotopes, except for a too high value of E(2$^{+}_{1})$ in $^{54}$Ca~\cite{step}. The same trend is observed with the overestimation of E(2$^{+}_{1})$ in the neutron-rich argon isotopes~\cite{step2,liu}. The interaction was slightly modified to improve the agreement with new experimental data. Fine tuning of the main ingredients is detailed in Refs.~\cite{utsu,step2}. With these modifications, it was possible to reproduce fairly well the variation of $E(2^{+}_{1}$) for the Ar isotopes~\cite{step2,cort,liu}, with a small increase from $^{48}$Ar to $^{50}$Ar and a larger value for $^{52}$Ar, but still underestimation of $^{46}$Ar at $N$ = 28. The wave function of the closed-shell $^{52}$Ca ground state was found to be dominated by the $\nu (p^{4}_{3/2})$ and $\nu (p^{3}_{3/2} p_{1/2})$ neutron configurations, while the ground state wave function of $^{50}$Ar was more mixed~\cite{step2}. The energy splitting in potassium isotopes was well reproduced by shell model calculations using another tuning defined in Ref.~\cite{sun}, the SDPF-MU$s$ interaction, with the restoration of the $Z = 16$ sub-shell gap in $^{51,53}$K isotopes. The degeneracy of $\pi 1s_{1/2}$ and $\pi 0d_{3/2}$ orbitals at $N=28$ is expected to favor collectivity in neutron-rich $Z = 14, 16, 18$ isotopes. The spectroscopic information for $^{47,49}$Cl obtained in the present work thus provides an important benchmark for the calculations. 

	The \textit{ab initio} VS-IMSRG~\cite{bogner14,stroberg16,stroberg19} was also used in the analysis of the experimental results. This approach resembles the standard shell model paradigm in that it decouples a valence-space Hamiltonian from the full-space problem via an approximate unitary transformation. In the IMSRG(2) approximation employed here, all induced operators are truncated at the two-body level, while three-body interactions between valence particles are taken into account with the ensemble normal ordering procedure~\cite{stroberg}. In the present application, the valence space is composed of the standard $sd$ orbitals for protons and $pf$ orbitals for neutrons on top of a $^{28}$O core. The initial interaction is the 1.8/2.0 (EM) interaction~\cite{hebeler11}, which combines the two-nucleon (NN) potential of Ref.~\cite{entem03} evolved via SRG techniques to $\lambda_\text{SRG} = 1.8 \text{ fm}^{-1}$ and three-nucleon (3N) interactions determined by a fit to the $^3$H binding energy and the $^4$He radius at that scale using a cutoff $\Lambda_\text{3N} = 2.0 \text{ fm}^{-1}$. While the NN potential is derived at next-to-next-to-next-to-leading order (N$^3$LO) in chiral EFT, the 3N part contains N$^2$LO operators. This interaction has been shown to provide an accurate description of the energies of ground- and excited- states in numerous applications across medium- and even heavy-mass nuclei~\cite{simonis, morris18, stroberg21,miyagi21}. In this framework, chlorine isotopes were first addressed in Ref.~\cite{simonis16}, where ground-state energies were computed for the whole isotopic chain. 
	
	In addition, full-space \textit{ab initio} calculations within the Gorkov self-consistent Green's function approach~\cite{soma11, soma14a} were performed. This technique has been recently applied to specific examples~\cite{leist, chen19, sun, mougeot20}, as well as systematic surveys~\cite{soma20a, soma20c} in the medium-mass region of the nuclear chart. One specificity of Green's function techniques is the ability to access the spectroscopy of odd-even nuclei~\cite{soma11}, which makes it particularly suited to the present case. Two different (NN+3N) interactions were used in this work. The first one, labeled NNLO$_{\text{sat}}$, builds on the set of operators appearing up to N$^2$LO in chiral EFT. It was introduced in Ref.~\cite{ekstrom15} with the main goal of correcting for the poor saturation properties and unsatisfactory description of nuclear radii yielded by previous chiral Hamiltonians. Indeed, it has been shown to reproduce experimental charge radii all the way up to the Nickel chain and mass $A$ = 132 for Xe isotopes~\cite{soma20a,art20}. In addition, it leads to a good description of all observables that crucially rely on a correct account of the nuclear size, such as the electric dipole response~\cite{raimondi19} or electron-nucleus scattering cross sections~\cite{barbieri19}.
	
	The second interaction, labeled NN+3N(lnl), was introduced more recently~\cite{soma20a}. It combines a N$^3$LO two-body force with N$^2$LO three-body operators. It has been shown to provide a very good description of (total and differential) ground-state energies and excitation spectra in the calcium region~\cite{soma20a, soma20c}. All GGF calculations were performed in a spherical harmonic-oscillator basis, including up to 14 major shells ($e_\text{max}=13$). Matrix elements of three-body operators were further restricted to $e_\text{3max}=16$. All energies presented in the following are converged in terms of many-body and model-space truncations up to a few percent errors (see Ref.~\cite{soma20a} for an extensive discussion).
	
	\maketitle
	\subsection{\texorpdfstring{$^{49}$Cl}{Lg}}
	
	The level schemes obtained for the low-lying states of $^{49}$Cl with the SDPF-MU and IMSRG calculations are shown in Fig.~\ref{fig5} and Tab.~\ref{tab2}. More compression is obtained in the IMSRG calculation with a lower-lying $1/2^{+}_{1}$ state, but a correspondence state-by-state may be done at similar excitation energies in both calculations. In contrast with the K isotopes, the calculation with the SDPF-MU$s$ interaction does not reproduce the experimental results well, with $J^{\pi}$ = 1/2$^{+}$ for the ground state of $^{49}$Cl. This is in contradiction with the positive $\Delta$ = 350 keV value found experimentally for $^{49}$Cl, much better reproduced here with the SDPF-MU interaction. Spectroscopic factors are shown in Tab.~\ref{tab2}. They suggest that the one-proton knockout reaction $^{50}$Ar(p,2p)$^{49}$Cl should significantly populate $1/2^{+}_{1}$ and $3/2^{+}_{1}$ states, exhausting most of the corresponding strength. The $0d_{5/2}$ strength seems to be more fragmented with a limited value for the spectroscopic factor of the $5/2^{+}_{2}$ state. This state may still be populated in the one-proton knockout. 
	
	The same picture, with a first excited state very close to the ground state, emerges in GGF calculations (see 
	Tab.~\ref{tab7}). However, the two interactions give different predictions for the ground-state spin, namely $3/2^+$ in the case of NNLO$_{\text{sat}}$ and $1/2^+$ for $NN$+$3N\text{(lnl)}$. In both cases, the energy of the first excited state is comparable with those found by shell model calculations, around 200 keV, slightly smaller than the experimental value. Interestingly, NNLO$_{\text{sat}}$ provides the correct spin, in contrast to what is found in the K isotopes (see discussion in Sec.~\ref{secspininv}). The rest of both spectra is characterized by higher-lying states above 1.5 MeV, mainly with spins of $3/2^+$ and $5/2^+$. The two excited states $3/2^{+}_{2}$ and $5/2^{+}_{1}$ found with the SDPF-MU Hamiltonian are of a highly collective nature and are not accounted for by the correlations currently included in GGF.
	\subsection{\texorpdfstring{$^{47}$Cl}{Lg}}
	
	In Tab.~\ref{tab7}, the energies and spectroscopic factors for low-lying states in $^{47}$Cl populated via the $^{48}$Ar(p,2p) channel obtained with SDPF-MU and IMSRG calculations are displayed. With a similar level scheme for both calculations, it is obvious that the $3/2^{+}_{1}$ and $1/2^{+}_{1}$ states have a strong overlap with proton-hole states in $^{48}$Ar. Similarities may also be observed with $^{49}$Cl in Tab.~\ref{tab2}.
	
	Due to the large momentum acceptance, other reaction channels populating bound states of $^{47}$Cl were analyzed. At first, the one-neutron knockout reaction $^{48}$Cl(p,pn)$^{47}$Cl is considered. The spin of the incoming $^{48}$Cl projectiles on the LH$_2$ target is not known. Three low-lying states of $^{48}$Cl, very close to each other, are predicted by the SPDF-MU calculation, the $J^{\pi}$ = 2$^{-}_{1} $ ground state, the 0$^{-}_{1}$ state at 78 keV, and the 1$^{-}_{1}$ state at 119 keV. From calculated $B(E2$) and $B(M1)$ values, isomerism is possible, but the decay times are small, compared to the 655 ns necessary to cross 118 meters between the primary and LH$_{2}$ target at v/c = 0.6. 
	
	Spectroscopic factors for low-lying states in $^{47}$Cl are displayed in Tab.~\ref{tab48Cl47Cl}, assuming the $^{48}$Cl projectile in one of the three possible states, $2^{-}_{1}$, $0^{-}_{1}$ or $1^{-}_{1}$. The cross sections $\sigma_{th}$ are calculated using Eq.~\ref{sigma} and the single-particle cross sections calculated with the TC method~\cite{moro}. The relative population of the $1/2^{+}_{1}$ vs $3/2^{+}_{1}$ states evolves with the spin of the $^{48}$Cl projectile, with a maximum value found for the $0^{-}_{1}$ state. In all cases, small cross-sections for the $1/2^{+}_{1}$ and $3/2^{+}_{1}$ states are obtained, compared to the experimental inclusive value $\sigma_{inc}$ = 19.3(15) mb. The difference is even larger for the $5/2^{+}_{1}$ or $3/2^{+}_{2}$ levels between the calculated cross-sections displayed in Tab.~\ref{tab48Cl47Cl} and the experimental value 12.7(15) mb obtained for the structure around 600 keV.
	
	\begin{figure}[hbtp!]
		\begin{center}
			\includegraphics[width=7.8 cm]{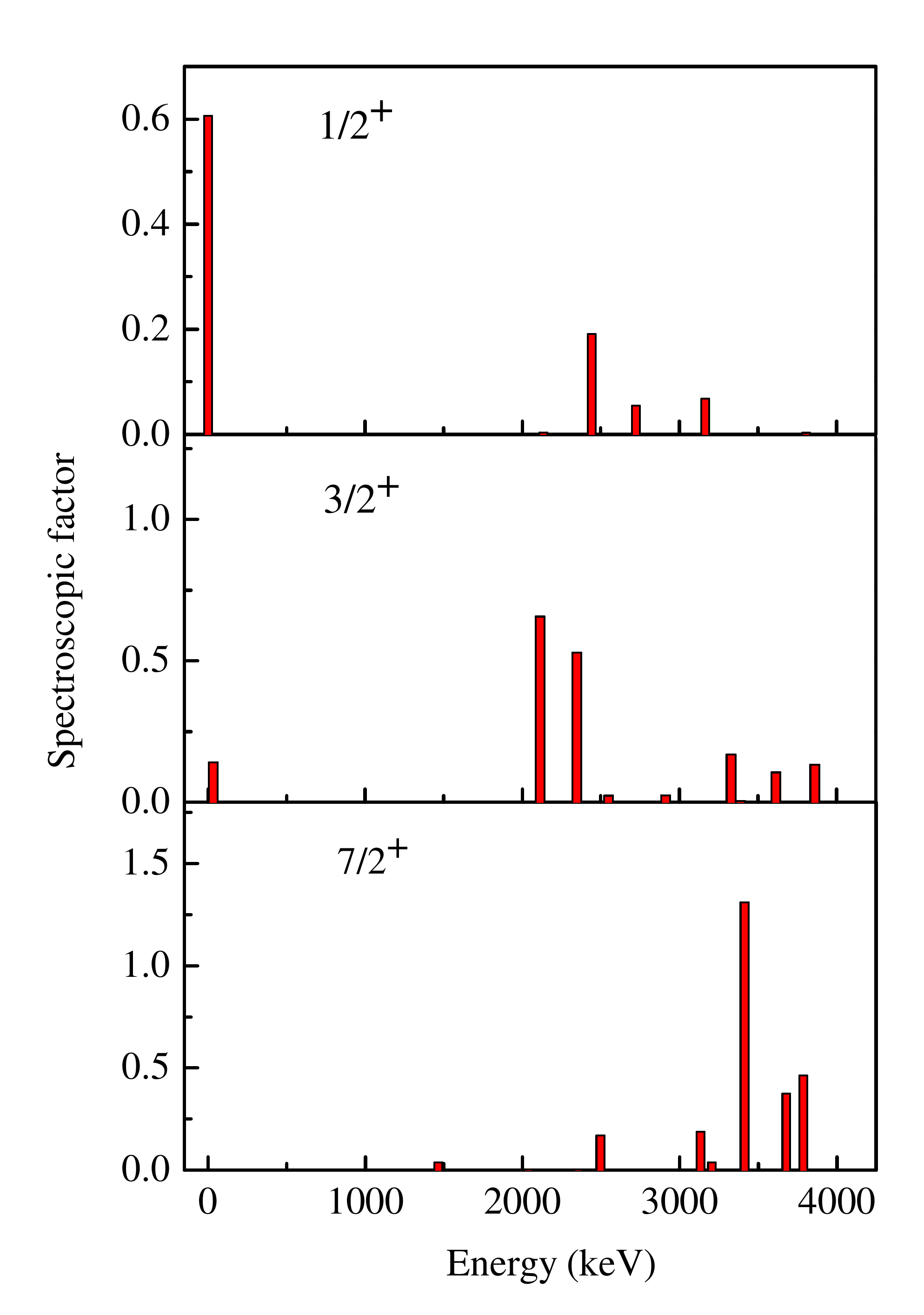}
		\end{center}
		\caption{Spectroscopic factor distributions of $^{47}$Cl states populated in the one neutron knockout reaction $^{48}$Cl(p,pn) calculated for different orbitals $n\ell j$, with the SDPF-MU interaction.$^{48}$Cl projectiles are taken in the 0$^{-}_{1}$ state.}
		\label{fig10}
	\end{figure}
	
	These differences can be understood by considering the population of higher lying states, as suggested in Fig.~\ref{fig10} from the SDPF-MU calculation in the simplest case of $^{48}$Cl in the $0^{-}_{1}$ state. 
	While the ground state is mainly populated in the removal of a $1p_{1/2}$ neutron, the distributions of the $3/2^{+}$ and $7/2^{+}$ states populated by $1p_{3/2}$ and $0f_{7/2}$ neutron removal are spread at higher excitation energies. 
	These high-lying states do not necessarily decay directly to the ground state. Depending on the level density at high excitation energy and possibly complex branching ratios, the low-lying 
	levels may be indirectly populated. This could explain the large cross section associated with the 600 keV structure, while the direct population of $5/2^{+}_{1}$ or $3/2^{+}_{2}$ by one-neutron removal is much smaller in Tab.~\ref{tab48Cl47Cl}.
	Contrarily, the SDPF-MU calculation predicts a strong direct population of the $3/2^{+}_{3}$ (2113 keV) and $3/2^{+}_{4}$ (2348 keV) states, with a large branching ratio for the direct decay to the ground state, consistent with a weak feeding of the first excited state.
	
	Therefore, the non-observation of the 148 keV transition in Fig.~\ref{fig7}-b) could be related to the selectivity of the one-neutron knockout reaction populating either the ground state or states lying at much higher excitation energy with specific decay-paths which do not feed the 148 keV state.

	The broad structure observed at 600 keV may have two different origins consistent with data, as shown in Fig.~\ref{fig9}: $i$) in the hypothesis (1), two states at 578 and 634 keV directly decay to the ground state with no - or very weak - decay towards the first excited state at 148 keV. Their excitation energies are compatible with the $5/2^{+}_{1}$ and $3/2^{+}_{2}$ predicted in the IMSRG or SDPF-MU calculations. From energies and $B(E2)$ and $B(M1)$ values obtained in the SDPF-MU calculation, a prompt decay to the ground state is predicted either for the $5/2^{+}_{1}$ or $3/2^{+}_{2}$ level, which supports the assumption of a doublet at 578 and 634 keV, rather than a single state at 624 keV with lifetime; $ii$) in the hypothesis (2), only one state at 634 keV decays either to the ground state or the first excited state, which are nearly degenerate in energy. The very small energy difference (56 keV) is consistent with the IMSRG and SDPF-MU calculations. In both calculations, two states have an excitation energy compatible with 634 keV, the $5/2^{+}_{1}$ or $3/2^{+}_{2}$ states, populated either directly or from higher-lying states. However, they are less compatible with the state at 148 keV. In order to disentangle the two hypotheses, the unambiguous identification of the first excited state is needed.

	In the case of the multi-nucleon removal reaction $^{50}$Ar(p,2p2n), the direct population of single-particle states in $^{47}$Cl is not expected to be strong in this non-direct mechanism, except for decay from higher-lying states. High spin states with np-nh configurations would be more favored, like the $7/2^{+}$, $9/2^{+}$, $11/2^{+}$ states present in the shell model calculations. The non-selectivity of the multi-nucleon removal process may explain the more complex decay paths resulting in different feedings of the low-lying states compared to the one-neutron knockout reaction and the presence of the 148 keV transition. 

	An open question remains for the energy of the first excited state in $^{47}$Cl and whether the transition at 148 keV originates from the decay of this state to the ground state or not. The excitation energy $E^{*}$ of this state predicted to be 3/2$^{+}_{1}$ in Tab.~\ref{tab7}, is below 200 keV in all calculations. If $E^{*}$ is above the experimental energy threshold around 100 keV, then the 148 keV transition is the best candidate for the transition between this state and the ground state. However, it may be well below the energy threshold as predicted in the SDPF-MU and IMSRG calculations. In this case, the energy threshold combined with the large atomic background will make it difficult to observe this transition in the experimental spectrum. Then, the origin of the strong 148 keV transition would be difficult to understand except for more compression in the calculated spectra, the 3/2$^{+}_{2}$ and 5/2$^{+}_{1}$ states being at higher excitation energies. 

	GGF approach could be performed for $^{47}$Cl only with the NNLO$_{\text{sat}}$	interaction, but not with the $NN$+$3N\text{(lnl)}$ interaction due to problems of convergence\footnote{As suggested by the IMSRG calculations, performed with very similar potential, the energies of the ground-state and the first excited state are presumably very close in energy for this particular interaction. The iterative character of the GGF solution (as opposed to the full diagonalization performed in valence-space calculations), the collective nature of these states, and the not-optimal account of quadrupole correlations~\cite{soma20c} make the resolution of this low-lying spectrum particularly challenging for the current GGF implementation.}. The first excited state is found at 210 keV with spin $1/2^{+}$, at variance with the results of other calculations displayed in Tab.~\ref{tab7}, although all calculations predict a first excited state at low energy, similar to $^{49}$Cl. The $5/2^{+}_{1}$ and $5/2^{+}_{2}$ states are predicted at higher energy, well above the values obtained by the SDPF-MU and IMSRG calculations displayed in Fig.~\ref{fig9}. Again, this can be ascribed to collective correlations missing in the GGF approach.
	
	To conclude on the level scheme and first excited state in $^{47}$Cl, a more specific experiment is needed, such as the one-proton knockout reaction $^{48}$Ar(p,2p) with measurement of the momentum distribution and an appropriate energy threshold.

	\begingroup
	\renewcommand*\arraystretch{1.1}
	\begin{center}
		\begin{table*}[hbtp!]
		\caption{Spin, excitation energies, and spectroscopic factors $ || \langle ^{A-1}Cl$ + p $\Vert ^{A}Ar \rangle || ^2 $ of low-lying states for $^{45,47,49}$Cl calculated with shell-model and ab initio approaches (see Section VI). }
		\begin{ruledtabular}
			\begin{tabular*}{0.8\textwidth}{@{\extracolsep{\fill}}cccccccccccccc}
				&& \multicolumn{2}{c}{SDPF-MU} &\multicolumn{2}{c}{SDPF-MU$_{}s$}
				&\multicolumn{2}{c}{IMSRG}
				&\multicolumn{2}{c}{GGF(NNLO$_{\text{sat}}$)}
				&\multicolumn{2}{c}{GGF($NN$+$3N\text{(lnl)}$)} \\
				\cmidrule(lr){3-4}
				\cmidrule(lr){5-6}\cmidrule(lr){7-8} \cmidrule(lr){9-10}\cmidrule(lr){11-12}
				Nucleus&State&Energy&$C^{2}S$&Energy&$C^{2}S$&Energy&$C^{2}S$&Energy&$C^{2}S$&Energy&$C^{2}S$\\
				&&(keV)&&(keV)&&(keV)&&(keV)&&(keV)&\\
				
				\midrule
				\multirow{2}{*}{$^{45}$Cl}&1/2$^{+}_{1}$& gs&0.77&gs&0.75& 14&0.67&110&1.51&gs&1.57\\
				&3/2$^{+}_{1}$&   174&2.62&264&2.54&gs&2.59&gs&1.21&140&1.62\\
				\midrule
				\multirow{2}{*}{$^{47}$Cl}&1/2$^{+}_{1}$&    gs&1.00&gs&0.94& gs&0.79&210&1.39&n.a&n.a\\
				&3/2$^{+}_{1}$&   33&1.85&193&1.96&8&2.29&gs&1.46&n.a&n.a\\
				\midrule
				\multirow{2}{*}{$^{49}$Cl}&3/2$^{+}_{1}$&   gs&1.91&83&2.03&gs&2.53&gs&1.48&160&1.44\\
				&1/2$^{+}_{1}$&  419&1.40&gs&1.27& 135&1.05&230&1.51&gs&1.49\\
			\end{tabular*}
			\end{ruledtabular}
			
		\label{tab7}
		\end{table*}
	\end{center}
	\endgroup
				
	\begingroup
	\renewcommand*\arraystretch{1.1}
	\begin{center}
		\begin{table*}[hbtp!]
    		\caption{Spin, excitation energies and spectroscopic factors calculated with the SDPF-MU interaction for the $^{47}$Cl states populated in the $^{48}$Cl(p,pn) reaction. Three different low-lying states, 0$^{-}_{1}$, 2$^{-}_{1}$ (ground state), and 1$^{-}_{1}$ are considered for the incident $^{48}$Cl projectiles. Single-particle cross sections $\sigma_{sp}^{lj}(E^{*},E_{inc})$ are calculated with the TC method~\cite{moro}.}
    	\begin{ruledtabular}
			\begin{tabular*}{0.8\textwidth}{@{\extracolsep{\fill}}ccccccccccccccccc}
				\multicolumn{2}{c}{$^{48}$Cl}&\multicolumn{2}{c}{0$^{-}_{1}$~78~keV}&\multicolumn{5}{c}{2$^{-}_{1}$~gs}&\multicolumn{5}{c}{1$^{-}_{1}$~119~keV}\\
				\cmidrule(lr){1-2} \cmidrule(lr){3-4} \cmidrule(lr){5-9} \cmidrule(lr){10-14}
				 State&E&$C^{2}S$&$\sigma_{th}$&\multicolumn{4}{c}{$C^{2}S$}&$\sigma_{th}$&\multicolumn{4}{c}{$C^{2}S$}&$\sigma_{th}$\\
			   \cmidrule(lr){1-2}\cmidrule(lr){3-4}\cmidrule(lr){5-8}\cmidrule(lr){10-13}
				$^{47}$Cl     &(keV)&&(mb)&0f$_{7/2}$&0f$_{5/2}$&1p$_{3/2}$&1p$_{1/2}$&(mb)&0f$_{7/2}$&0f$_{5/2}$&1p$_{3/2}$&1p$_{1/2}$&(mb)\\
				\midrule
				1/2$^{+}_{1}$&gs &0.606  &5.2&&0.005&0.064& &0.5&&&0.152&0&1.2\\
				3/2$^{+}_{1}$&33&0.141   &1.1&0.023&0.005&0.213&0.099&2.7&&0&0.018&0.541&4.8\\
				5/2$^{+}_{1}$&714&0.007  &0.04&0.148&0&0.005&0.024&1.&0.029&0&0.006&&0.2\\
				3/2$^{+}_{2}$&785&0      &0  &0.06& 0.01&0.058&0.005&0.9&&0.002&0.003&0.013&0.1\\
				\midrule
				3/2$^{+}_{3}$&2113 &0.657  &4.4&&&& &&&&&&\\
				3/2$^{+}_{4}$&2348 &0.529  &3.5&&&& &&&&&&\\
				7/2$^{+}_{7}$&3413 &1.131  &6.8&&&& &&&&&&\\
			\end{tabular*}
		\end{ruledtabular}
			\label{tab48Cl47Cl}
		\end{table*}
	\end{center}
	\endgroup
	\vspace{-1.5 cm} 

	\subsection{Spin inversion over K and Cl isotopic series}
	\label{secspininv} 

	The evolution of the energy difference $\Delta = E(1/2^{+}_{1}) - E(3/2^{+}_{1})$ with increasing neutron number N has been investigated for K isotopes~\cite{papu1,papu2} and recently extended up to $N = 34$~\cite{sun}. In the context of the shell model~\cite{sun}, it was proposed that the decrease of $\Delta$ from $N = 20$ to 28, roughly linear with $N$, is due to the attractive interaction $\pi d_{3/2} - \nu f_{7/2}$ with a maximum effect at $N = 28$ and an associated spin inversion. Analogously, the opposite effect beyond $N = 28$ and the spin re-inversion at $N = 32$ are ascribed to the filling of the $\nu p_{3/2}$ orbital and the $\pi s_{1/2} - \nu p_{3/2}$ interaction, which is more attractive than the $\pi d_{3/2} - \nu p_{3/2}$ one. 
	
\begin{figure}[htbp]
		\begin{center}
			\includegraphics[width=8.6cm]{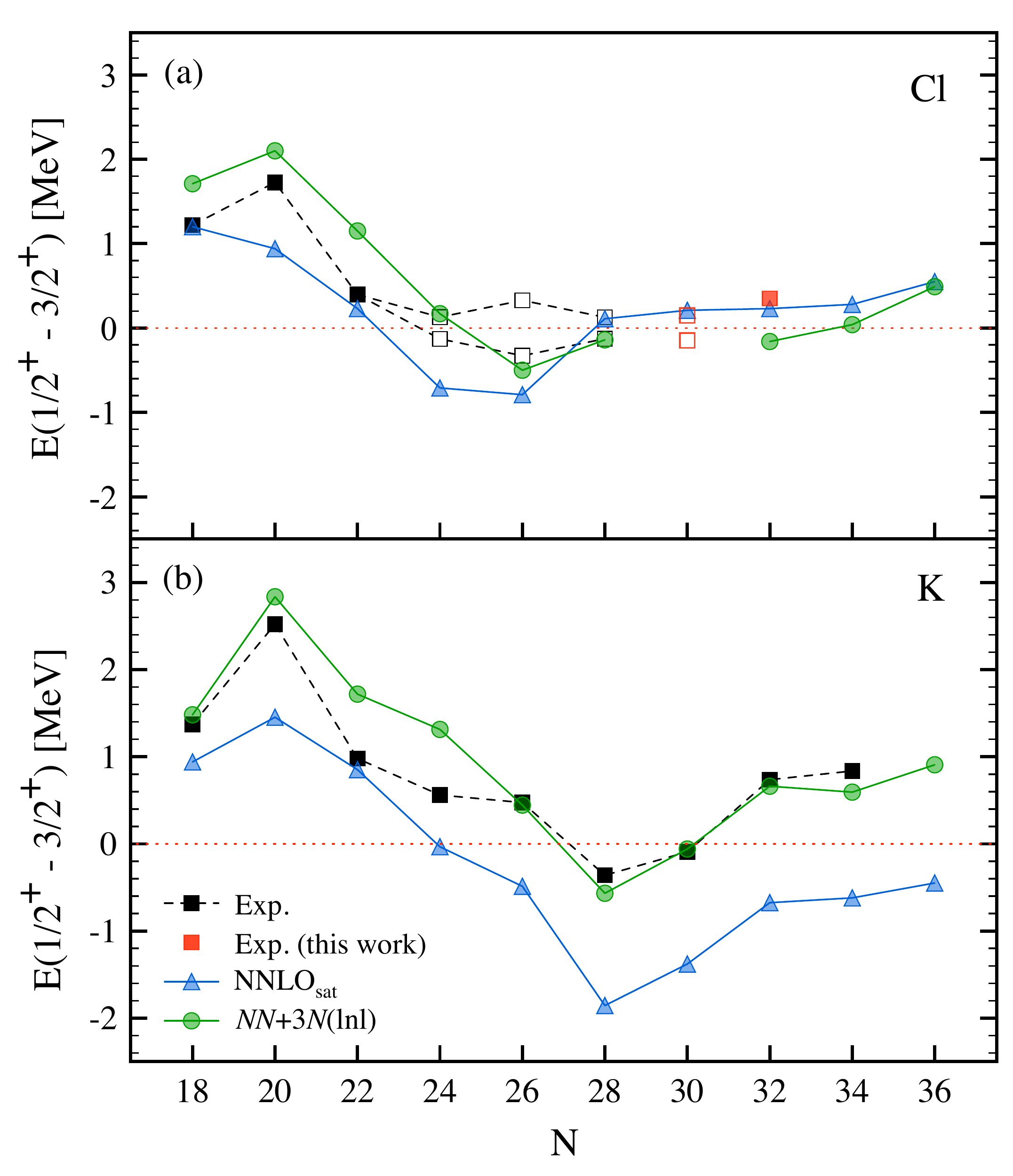}
		\end{center}
		\caption{Variation of the energy difference between the first $1/2^{+}$ and $3/2^{+}$ states for Cl and K isotopes with the neutron number $N$. GGF calculations were performed with NNLO$_{\text{sat}}$ and $NN$+$3N\text{(lnl)}$ interactions. Full squares are experimental values. When only the absolute value of the energy difference $\Delta$ has been established, empty squares are added following a $1/2^{+}_{1}$ or $3/2^{+}_{1}$ hypothesis for the ground state. No value is given for the calculation of $^{47}$Cl with the $NN$+$3N\text{(lnl)}$ interaction, as explained in text. Experimental values for $^{41,43,45}$Cl are taken from~\cite{gad,lia,oll,szi,wiz,ibb,sor}; experimental values for K isotopes are taken from~\cite{papu1, papu2,sun}.
		}
		\label{fig11} 
	\end{figure}

	\textit{Ab initio} GGF calculations were able to provide very good reproduction of the evolution of $\Delta$ along K isotopes~\cite{sun, soma20a}, see Fig.~\ref{fig11}-(b). In particular, the $NN$+$3N\text{(lnl)}$ Hamiltonian yields a remarkable agreement with the experimental energy difference throughout the whole chain and correctly reproduces the spin inversion and re-inversion. The NNLO$_{\text{sat}}$ Hamiltonian, in contrast, provides a good general description of the experimental trend but a $\Delta$ that is too low in absolute energy, thus failing to describe the re-inversion at $^{51}$K.

	In this work, GGF calculations of $\Delta$ were extended to Cl isotopes from $N = 18$ to $N= 36$, see Fig.~\ref{fig11}-(a). Around the valley of stability, a behavior analogous to the K case is observed with a maximum of $\Delta$ at $N = 20$ and a subsequent decrease when neutrons are successively added on the $\nu f_{7/2}$ orbital. The two interactions provide similar trends with NNLO$_{\text{sat}}$ shifted down roughly by 1 MeV with respect to $NN$+$3N\text{(lnl)}$. The former (latter) predicts a spin inversion of the ground state at $N = 24$ ($N = 26$). Experimentally, while for $N = 24$, 26, and 28 isotopes the energy of the first excited state is known, its spin (and that of the ground state) could not be firmly established, such that only the absolute value of $\Delta$ is available. Hence, none of the possibilities can be ruled out. Beyond $N = 26$, a change in behavior is observed. Both available experimental data and theoretical curves remain close to zero all the way up to $N = 34$. In particular, the regular trend characterizing K isotopes, with a monotonic decrease and a subsequent increase past $N = 28$, is not reproduced here. By inspecting the theoretical strength distribution, we notice that states around the Fermi surface in Cl isotopes are much more fragmented than in their neighboring $Z + 2$ isotones. The partial occupations of $1/2^+$ and $3/2^+$ states thus likely washes out the mechanism at play in the K isotopes, which relies on ``naive'' occupations of $\pi s_{1/2}$ and $\pi d_{3/2}$ shells. Two different aspects induce this behavior. First, simply two fewer protons are available for the $\pi d_{3/2} - \nu f_{7/2}$ interaction to operate, with a subsequent weakening of the attractive effect. Second, several indications suggest that Ar nuclei around $N = 28$ constitute a transitional region between spherical Ca and deformed S isotopes (see e.g. a recent discussion in Ref.~\cite{mougeot20}). In particular, $^{44, 46, 48}$S are thought to be characterized by static deformation with either oblate or prolate minima~\cite{wer,guz,sohl}. This picture is indeed consistent with a reduced $Z = 16$ gap and a mixing of configurations involving the $\pi s_{1/2}$ and $\pi d_{3/2}$ orbitals, with a subsequent fragmentation around the Fermi surface. These features make the study of Cl isotopes more challenging for theoretical approaches. Future measurements aiming at pinning down the sign of $\Delta$ between $N = 24$ and $N= 30$ will provide a unique test-bench for the development of both shell model interactions and \textit{ab initio} methods.

	\section{Conclusion}
	
	To summarize, spectroscopy of the neutron rich $^{47,49}$Cl isotopes at $N = 30, 32$ was carried out for the first time. The main reaction was the one-proton knockout $^{50}$Ar(p,2p)$^{49}$Cl with detection of photons emitted in-flight, coupled to the measurement of the momentum distributions of the residues. Due to the large acceptance of the SAMURAI spectrometer, multi-nucleon removal reactions were also analyzed. The ground state of $^{49}$Cl was found to be consistent with a $J^{\pi} = 3/2^{+}$ assignment and a $1/2^{+}$ first excited state. This normal ordering for $3/2^{+}$ and $1/2^{+}$ states is similar to the recently observed $^{51}$K case, while spin inversion is still under debate for the less neutron-rich chlorine isotopes $^{41,43,45,47}$Cl.

	\begin{acknowledgments}
	
	We thank the RIKEN Nishina Center accelerator staff for their work in the primary beam delivery and the BigRIPS team for preparing the secondary beams. 
	The development of MINOS has been supported by the European Research Council through the ERC Grant No. MINOS258567. 
	B. D. L., L. X. C., and N. D. T. acknowledge support from the Vietnam Ministry of Science and Technology under Grant No. {\DJ}TCB.01/21/VKHKTHN. F.B. was supported by the RIKEN Special Postdoctoral Researcher Program. 
	Y. L. S. acknowledges the support of Marie Sk\l{}odowska-Curie Individual Fellowship (H2020-MSCAIF-2015-705023) from the European Union. 
	I. G. has been supported by HIC for FAIR and Croatian Science Foundation.
	R. -B. G. is supported by the Deutsche Forschungsgemeinschaft (DFG) under Grant No. BL 1513/1-1.
	K. I. H., D. K. and S. Y. P. acknowledge the support from the IBS grant funded by the Korea government (No. IBS-R031-D1).
	P. K. was supported in part by the BMBF grant No. 05P19RDFN1 and HGS-HIRe.
	D. So. has been supported by the European Regional Development Fund contract No. GINOP-2.3.3-15-2016-00034 and the National Research, Development and Innovation Fund of Hungary via Project No. K128947.
	This work was supported in part by JSPS KAKENHI Grants No. JP16H02179, JP18H05404 and JP20K03981. 
	J. D. H. and R. S. acknowledge the support from NSERC and the National Research Council Canada. 
	This work was supported by the Office of Nuclear Physics, U.S. Department of Energy, under Grants No. de-sc0018223 (NUCLEI SciDAC-4 collaboration) and the FieldWork Proposal ERKBP72 at Oak Ridge National Laboratory (ORNL). Computer time was provided by the Innovative and Novel Computational Impact on Theory and Experiment (INCITE) program. This research used resources of the Oak Ridge Leadership Computing Facility located at ORNL, which is supported by the Office of Science of the Department of Energy under Contract No. DE-AC05-00OR22725.
	GGF calculations were performed by using HPC resources from GENCI-TGCC (Contracts No. A007057392, A009057392) and at the DiRAC Complexity system at the University of Leicester (BIS National E-infrastructure capital grant No. ST/K000373/1 and STFC grant No. ST/K0003259/1). This work was supported by the United Kingdom Science and Technology Facilities Council (STFC) under Grant No. ST/L005816/1 and in part by the NSERC Grant Nos. SAPIN-2016-00033, SAPIN-2018-00027, and RGPAS-2018-522453. TRIUMF receives federal funding via a contribution agreement with the National Research Council of Canada. 
	J. D. H. thanks S. R. Stroberg for the imsrg++ code used to perform the VS-IMSRG calculations https://github.com/ragnarstroberg/imsrg.
	N.T.T.P. was funded by Vingroup Joint Stock Company and supported by the Domestic PhD Scholarship Programme of Vingroup Innovation Foundation (VINIF), Vingroup Big Data Institute (VINBIGDATA), code VINIF.2020.TS.52.
	\end{acknowledgments}
	
	\bibliography{references}

\end{document}